\documentclass[aps,prd,twocolumn,superscriptaddress,bibnotes,longbibliography,preprintnumbers,floatfix,10pt]{revtex4-2}

\usepackage{graphicx}
\usepackage{xcolor}
\usepackage{amsmath,amsfonts,amsthm,amssymb}
\usepackage[colorlinks]{hyperref}
\usepackage{mathtools}
\usepackage{bm}
\usepackage{multirow}
\usepackage{booktabs}
\usepackage{rotating}

\hypersetup{
    pdfnewwindow=true,
    colorlinks=true,
    linkcolor=violet,
    citecolor=violet,
    filecolor=violet,
    urlcolor=violet
}

\graphicspath{{./figures/}}

\newcommand{\orcid}[1]{\href{https://orcid.org/#1}{\includegraphics[width=10pt]{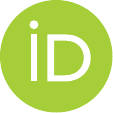}}}

\newcommand{\Cqed}{\mathcal{C}}

\newcommand{\dd}{\mathrm{d}}

\begin{document}

\title{Estimating High-Energy Neutrino Effective Areas from Muon Propagation}

\author{Stephan A.\ Meighen-Berger \orcid{0000-0001-6579-2000}\,}
\email{stephan-meighen-berger@uiowa.edu}
\affiliation{\href{https://ror.org/036jqmy94}{University of Iowa}, Iowa City, Iowa 52242, USA}

\date{\today}

\begin{abstract}
Neutrino telescopes observe neutrinos through the muons produced by charged-current interactions. Most of those muons are born kilometers outside the detector, and how much energy they lose on the way is stochastic. We show that we can compute analytically the distance a muon travels before it falls below the detection threshold, while describing the statistical properties of its losses. Our result reproduces Monte Carlo propagation to within $3\%$, and with two parameters per detector it matches the effective areas published by IceCube, KM3NeT/ARCA, and P-ONE to about $1\%$. We then show how this model can be used in point-source sensitivities, energy reconstruction, and consistency checks between detectors. The calculation disentangles muon-transport physics from a detector's response, so a new detector or loss model can be quickly benchmarked.
\end{abstract}

\maketitle

%%%%%%%%%%%%%%%%%%%%%%%%%%%%%%%%%%%%%%%%%%%%%%%%%%%%%%%%%%%%%%%%%%%%%%%%%%%%%%%%%%%%%%%%%%%
%%%%%%%%%%%%%%%%%%%%%%%%%%%%%%%%%%%%%%%%%%%%%%%%%%%%%%%%%%%%%%%%%%%%%%%%%%%%%%%%%%%%%%%%%%%

\section{Introduction}
\label{sec:intro}

%%%%%%%%%%%%%%%%%%%%%%%%%%%%%%%%%%%%%%%%%%%%%%%%%%%%%%
\begin{figure}[t]
\centering
\includegraphics[width=\columnwidth]{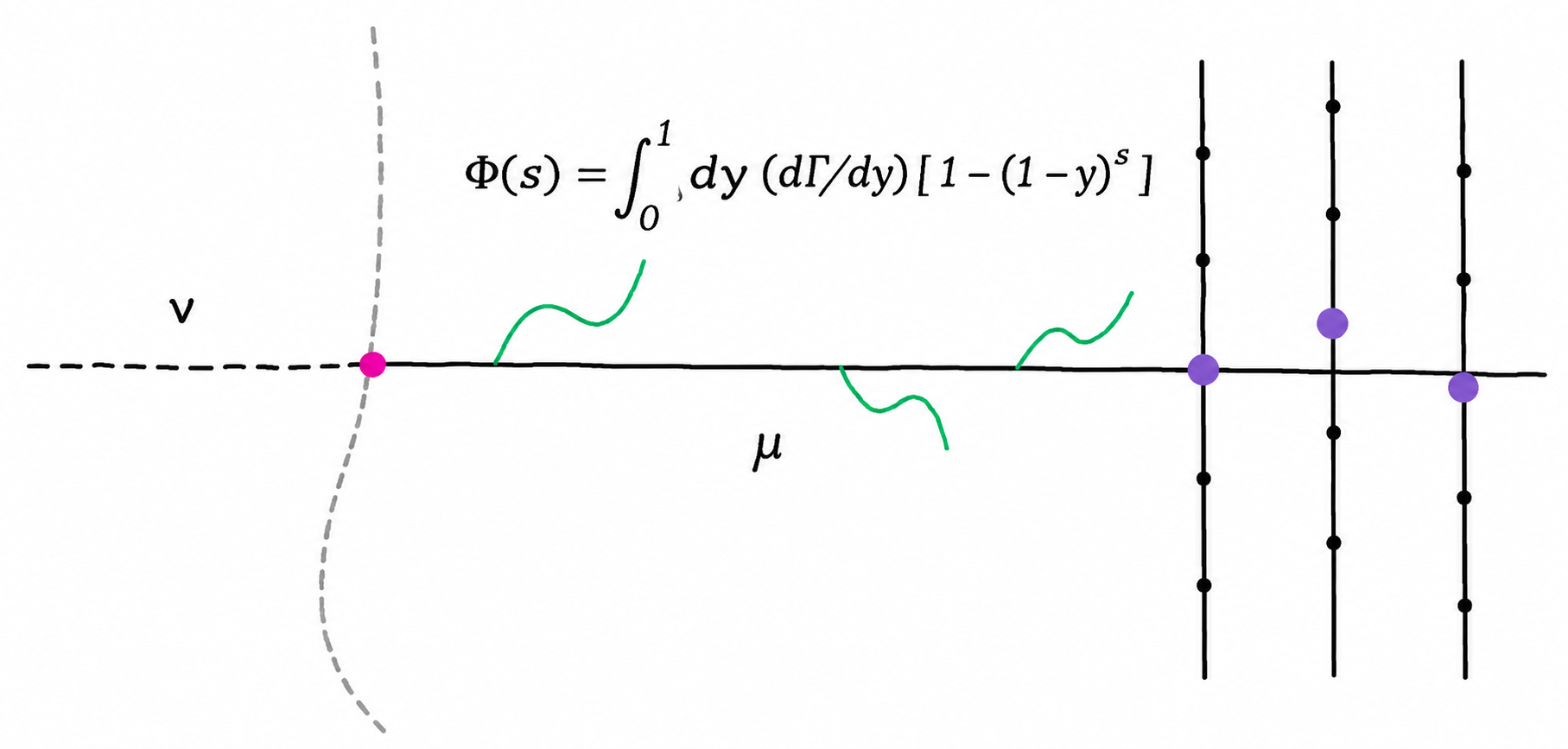}
\caption{A sketch of our approach. A neutrino produces a muon outside of the detector, and our model then follows the loss history of the muon through one function, $\Phi$, before it reaches the detector.}
\label{fig:sketch}
\end{figure}
%%%%%%%%%%%%%%%%%%%%%%%%%%%%%%%%%%%%%%%%%%%%%%%%%%%%%%

High-energy neutrino telescopes instrument large volumes of ice or water, up to a cubic kilometer, with light sensors~\cite{ANTARES:2011hfw, KM3Net:2016zxf, IceCube:2016zyt, Baikal-GVD:2022fis}. A neutrino that interacts in or near that volume produces a charged lepton, and the sensors record the Cherenkov light it emits along its path.

Over the past decade this program has established a diffuse astrophysical flux~\cite{IceCube:2013low, IceCube:2016umi, IceCube:2020wum} and resolved the first steady point sources~\cite{IceCube:2016cqr, IceCube:2018dnn, IceCube:2022der, IceCube:2023ame}. KM3NeT has since recorded a muon of about $120$~PeV, from possibly the most energetic neutrino observed so far~\cite{KM3NeT:2025npi}. Larger arrays are now being built or planned at new sites, in the Pacific, in the South China Sea and at the South Pole~\cite{P-ONE:2020ljt, TRIDENT:2022hql, IceCube-Gen2:2020qha}.

Most of what these telescopes record are muons. A muon neutrino that interacts through the charged current produces a muon, and at high energies that muon can travel many kilometers. It leaves a long track of light that points back to its source, so point-source searches are built on tracks.

The number of tracks a telescope records per unit time from a neutrino flux is, to first approximation
\begin{equation}
  \frac{\dd N}{\dd t} = \int \dd E_\nu \, \phi_\nu(E_\nu) \, V \, n_N \, \sigma_{\mathrm{CC}}(E_\nu) \, \epsilon,
  \label{eq:estimate}
\end{equation}
where $\phi_\nu$ is the neutrino flux, $V$ the effective volume, $n_N$ the number density of target nucleons, $\sigma_{\mathrm{CC}}$ the charged-current cross section, and $\epsilon$ the efficiency with which the detector records and selects a track. The product $n_N \sigma_{\mathrm{CC}}$ is the probability per unit length that a neutrino interacts, so a flux crossing the volume $V$ produces $\phi_\nu n_N \sigma_{\mathrm{CC}} V$ interactions per unit time and energy. The detector records a fraction $\epsilon$ of them. Typically everything but the flux is grouped into one object, the effective area~\cite{Gaisser:2016uoy}. Conservatively, we can set $V$ to the instrumented volume. But \textit{why is this conservative?}

A muon-neutrino charged-current interaction hands part of its energy to a muon, and the muon is counted if it reaches the detector above a threshold $E_{\mathrm{thr}}$. The effective volume is therefore the instrument's projected area times the distance over which a muon can be born and still be counted, which is its muon range. For a $1$~PeV muon and a TeV threshold the muon range is $\sim15$~km, so with IceCube's kilometer-scale footprint~\cite{IceCube:2016zyt} the effective volume is of order $15$~km$^3$, ten times the instrumented one.
\textit{Most of the sample is made outside the detector, so an effective area is both a transport and a detector question.}

Typically we calculate the volume in one of two ways. A Monte Carlo propagator samples the losses collision by collision~\cite{Antonioli:1997qw, Sokalski:2000nb, Chirkin:2004hz, Kudryavtsev:2008qh, Koehne:2013gpa, Dunsch:2018nsc, Safa:2021ghs, Garg:2022ugd, NuSpaceSim:2023ims}, and the cost of that sampling sets the pace of every detector study~\cite{Chirkin:2013tma, Lazar:2023rol, IceCube:2020tcq}. A design study that moves a string or lowers a threshold waits for a new simulation before it learns what the change bought. The other is a published instrument response, where the muon range is folded with the selection efficiencies of one analysis~\cite{IceCube:2019cia, Abbasi:2026ehs}. While this approach is fast, it is bound to the analysis and the detector that made it. It cannot be carried to a site that has not been built or to a question its analysis did not ask, and it cannot put two detectors on one footing.

Two recent papers proposed an analytic solution for the muon range~\cite{Palmisano:2025abd, Palmisano:2026sid}. They treat muon propagation as a stationary Boltzmann problem and approximate the solution by expanding the collision operator in the fraction of energy each collision removes. They leave a full solution to future work~\cite{Palmisano:2025abd}.

While the approximate solution captures the mean energy loss well, it stops too early, because muon transport is \textit{statistical}. In the IceCube IceTracks-DR2 release~\cite{Abbasi:2026ehs} the median reconstructed muon energy sits $1.6$ to $3.8$ decades below the parent. Fluctuations around that median can be as large as the median itself, since one hard bremsstrahlung photon may take most of the muon's energy at once~\cite{Lipari:1991ut, Kelner:1995hu, Dutta:2000hh, Safa:2021ghs}.

In this paper we take a different route and never solve for the energy distribution of the muon. A detector threshold sits at a fixed factor $E_0/E_{\mathrm{thr}}$ below the production energy $E_0$, so reaching it means losing a fixed amount in the logarithm of the energy. The muon range therefore needs only the statistics of that logarithmic loss: how fast it grows on average, and how much it scatters.

Both follow from one function, $\Phi$, once we treat the radiative losses as energy-scale invariant, meaning that the fraction of its energy a muon loses in a collision does not depend on the energy it has. The function $\Phi$ can be equated to the radiative loss coefficient $b_\mu$, the rate at which the mean energy of a muon falls, promoted from a number to a function. One of its values is $b_\mu$ itself, and its first two derivatives are the mean and the spread of the logarithmic loss, so the muon range follows from two derivatives of $\Phi$. At a fixed argument $\Phi$ is the Z~moment of cosmic-ray physics~\cite{Rossi:1941zza, Lipari:1993hd, Gaisser:2016uoy}, and we use it as a function.

The function $\Phi$ also needs the collisions only through a few integrals of their loss spectrum. We ask which of those integrals each quantity needs and use them, so we never need a unique reconstruction of the physical loss spectrum.

Figure~\ref{fig:sketch} sketches the idea. With the resulting muon range, the textbook estimate of an effective area becomes predictive. With nothing fitted it reproduces the published areas of four telescopes to about $10\%$, and three corrections take it to the percent level.

We structure this paper as follows: in Section~\ref{sec:exponent} we ask how far a muon travels, derive $\Phi$ and the muon range, and test both against the Monte Carlo propagator PROPOSAL. Section~\ref{sec:volume} turns the muon range into an effective area a reader can evaluate from a detector's geometry and two numbers, and Section~\ref{sec:theresponse} adds the three effects the estimate leaves out. Section~\ref{sec:aeff} holds both against four published effective areas and the IceCube event sample. Section~\ref{sec:applications} then turns the response on three problems where a simulation per point is expensive: the point-source sensitivity of four sites, the energy of one event, and a consistency check between detectors.

%%%%%%%%%%%%%%%%%%%%%%%%%%%%%%%%%%%%%%%%%%%%%%%%%%%%%%%%%%%%%%%%%%%%%%%%%%%%%%%%%%%%%%%%%%%
%%%%%%%%%%%%%%%%%%%%%%%%%%%%%%%%%%%%%%%%%%%%%%%%%%%%%%%%%%%%%%%%%%%%%%%%%%%%%%%%%%%%%%%%%%%

\section{How Far Does a Muon Travel?}
\label{sec:exponent}

The energy loss of a high-energy muon is usually written as
\begin{equation}
  -\frac{\dd E}{\dd \ell} = a_\mu(E) + b_\mu(E)\, E,
  \label{eq:dedx}
\end{equation}
with $\ell$ the column depth, the path length weighted by the density of the medium, $a_\mu$ the ionization loss and $b_\mu$ the radiative loss from bremsstrahlung, pair production and photonuclear collisions~\cite{Lohmann:1985qg, Groom:2001kq}. The radiative term takes over above the critical energy $E_c = a_\mu / b_\mu \simeq 600$~GeV in water.

The radiative term is a sum of discrete collisions. In each one the muon loses a fraction $y$ of its energy, $E \to (1-y)E$, with $0 < y < 1$. We describe these collisions by their loss spectrum $\dd\Gamma(E, y)/\dd y$. A muon of energy $E$ undergoes $(\dd\Gamma/\dd y)\,\dd y$ collisions with a fractional loss between $y$ and $y + \dd y$ per unit column depth. The coefficient $b_\mu$ is the first moment of that spectrum
\begin{equation}
  b_\mu(E) = \int_0^1 \dd y \, y \, \frac{\dd\Gamma(E, y)}{\dd y},
  \label{eq:bdef}
\end{equation}
the average fraction lost per unit column depth. We write $\langle f \rangle \equiv \int_0^1 \dd y \, f(y) \, \dd\Gamma/\dd y$ for such an integral over the loss spectrum, so $b_\mu = \langle y \rangle$. It is weighted by the collision rate and carries units of inverse column depth, unlike the mean inelasticities of Section~\ref{sec:volume}, which are plain averages.

Above $E_c$, integrating Eq.~(\ref{eq:dedx}) from a production energy $E_0$ down to a threshold $E_{\mathrm{thr}}$ gives the muon range
\begin{equation}
    R = \frac{1}{b_\mu}\ln (E_0/E_\mathrm{thr}),
    \label{eq:range_standard}
\end{equation}
and that range underlies the effective area of every neutrino telescope. However, it only describes the mean. A single muon's energy after a kilometer is one draw from a distribution with a long tail~\cite{Lipari:1991ut, Dutta:2000hh, Safa:2021ghs}, because one hard collision can take most of it at once.

How far, then, does a single muon travel before it first falls below $E_{\mathrm{thr}}$, and with what spread? A collision multiplies the energy by $(1-y)$, so it is natural to count the losses in the logarithm of the energy. In those units the threshold lies a fixed distance $w = \ln(E_0/E_{\mathrm{thr}})$ below the start, and the muon range is the column depth at which the accumulated logarithmic loss first exceeds $w$.

We will not solve for the distribution of the muon's energy itself. We derive the statistics of the logarithmic loss from one function, $\Phi$, in Section~\ref{sec:loglossW}, connect it to the transport equation of Refs.~\cite{Palmisano:2025abd, Palmisano:2026sid} in Section~\ref{sec:transporteq}, and evaluate it in Section~\ref{sec:collisions}.

\subsection{The Accumulated Logarithmic Loss}
\label{sec:loglossW}

At high energies the radiative losses are approximately energy-scale invariant~\cite{Lohmann:1985qg, Groom:2001kq}: the fraction of its energy a muon loses in a collision barely depends on the energy it has. Between $1$ and $100$~PeV in water, $b_\mu$ rises by only $15\%$~\cite{Koehne:2013gpa, Dunsch:2018nsc}. We assume the same of the whole spectrum
\begin{equation}
  \frac{\dd \Gamma(E, y)}{\dd y} = \frac{\dd \Gamma(y)}{\dd y}.
  \label{eq:scaleinv}
\end{equation}
We derive the transport with the spectrum held fixed, and then evaluate its integrals at the energy the muon has at each point along the track. The integrals, $b_\mu$ and the two the muon range needs below, thereby become slowly varying functions of energy (Table~\ref{tab:renewal}), and Appendix~\ref{app:renewal} integrates the muon range over them.

Ionization is not part of the loss spectrum. It removes a fixed amount of energy per unit column depth, unlike collisions, which remove fractions. Above $E_c$, it is only a small correction, and in Appendix~\ref{app:renewal} we show how to include it.

A collision that removes a fraction $y$ lowers the logarithm of the energy by $-\ln(1-y) > 0$. Summed along the track these steps give the accumulated logarithmic loss $W(\ell)$, and the muon's energy after a column depth $\ell$ is
\begin{equation}
  E = E_0 \, e^{-W(\ell)}.
  \label{eq:Wdef}
\end{equation}
Under Eq.~(\ref{eq:scaleinv}) the steps are independent and drawn from the same spectrum at every energy, so $W$ grows like a random walk that only moves forward.

The muon range will need two numbers, the mean of $W$ per unit column depth and its variance. The shortest way to both is to follow the average of $e^{-sW}$, where $s$ is an arbitrary variable. In a short column depth $\dd\ell$ a collision with fractional loss $y$ happens with probability $(\dd\Gamma/\dd y)\,\dd y\,\dd\ell$ and multiplies $e^{-sW}$ by $(1-y)^s$. Summed over all $y$, the average therefore falls at the rate
\begin{equation}
    \Phi(s) = \int_0^1 \dd y \, \frac{\dd \Gamma}{\dd y} \left[ 1 - (1-y)^s \right]
    \label{eq:Phi}
\end{equation}
per unit column depth, $\dd\langle e^{-sW}\rangle/\dd\ell = -\Phi(s)\,\langle e^{-sW}\rangle$, and since $W(0) = 0$ its solution is
\begin{equation}
  \langle e^{-sW} \rangle(\ell) = e^{-\ell \Phi(s)}.
  \label{eq:genfn}
\end{equation}

We keep $s$ continuous so that one function describes all the statistics of $W$ at once. 
% In the language of probability, $\Phi$ is the cumulant generating function of the logarithmic loss per unit column depth, up to signs. 
Differentiating the logarithm of Eq.~(\ref{eq:genfn}) at $s = 0$ returns the mean and the variance of $W$, so the first two derivatives of $\Phi$ are the mean and the variance of the logarithmic loss per unit column depth
\begin{eqnarray}
  \Phi'(0) &=& \int_0^1 \dd y \, \frac{\dd \Gamma}{\dd y} \left[ -\ln(1-y) \right], \nonumber \\
  -\Phi''(0) &=& \int_0^1 \dd y \, \frac{\dd \Gamma}{\dd y} \ln^2(1-y).
  \label{eq:Phimoments}
\end{eqnarray}
These two numbers are all the muon range will need (Section~\ref{sec:range}).

Setting $e^{-sW} = (E/E_0)^s$ turns Eq.~(\ref{eq:genfn}) into a statement about powers of the energy
\begin{equation}
  \langle E^s \rangle(\ell) = E_0^s \, e^{-\ell \Phi(s)}.
  \label{eq:momentlaw}
\end{equation}
By Eq.~(\ref{eq:bdef}) $\Phi(1) = b_\mu$, so the mean energy of a muon falls as $E_0 e^{-b_\mu \ell}$, the solution of Eq.~(\ref{eq:dedx}) above $E_c$. The other values of $s$ are new, and $\Phi(s)$ carries them all. It is $b_\mu$ promoted from a number to a function.

A few properties of $\Phi$ hold for any positive loss spectrum. It needs no infrared regulator, and it satisfies $\Phi(0) = 0$, $\Phi' > 0$ and $\Phi'' < 0$ (Appendix~\ref{app:diag}). A concave function through the origin has $\Phi'(0) > \Phi(1) = b_\mu$. The typical muon therefore descends faster than the mean energy does, because the mean is held up by the few muons that escaped a hard collision, and Section~\ref{sec:range} builds the muon range on that difference.

\subsection{The Transport Equation}
\label{sec:transporteq}

Refs.~\cite{Palmisano:2025abd, Palmisano:2026sid} start from a transport equation for the muon flux, and Eq.~(\ref{eq:momentlaw}) solves it. At ultra-relativistic energies the muon flux $\phi(\ell, E)$ is one-dimensional and obeys the gain-loss equation
\begin{equation}
  \frac{\partial \phi}{\partial \ell}
  = \Cqed[E, \phi] + S(\ell, E).
  \label{eq:transport}
\end{equation}
The source $S(\ell, E)$ injects muons, and in a telescope it is the charged-current interaction of the neutrino flux. For now we can ignore it. We follow a single muon, so we set $S = 0$ and start from $\phi(0, E) = \delta(E - E_0)$, which makes $\phi(\ell, E)$ the probability distribution of the muon's energy after a column depth $\ell$. We restore the source in Appendix~\ref{app:diag}. The first term is the collision operator
\begin{eqnarray}
  \Cqed[E, \phi] & = & - \phi(\ell, E) \int_0^1 \dd y \, \frac{\dd\Gamma(E, y)}{\dd y} \nonumber \\
  && + \int_0^1 \frac{\dd y}{1-y} \, \frac{\dd\Gamma(E_y, y)}{\dd y} \, \phi(\ell, E_y).
  \label{eq:collision}
\end{eqnarray}
Muons leave energy $E$ in every collision and arrive from $E_y = E/(1-y)$ above it, with $1/(1-y)$ the Jacobian of the shift.

To connect Eq.~(\ref{eq:transport}) with Eq.~(\ref{eq:momentlaw}), we multiply it by $E^s$ and integrate over $E$. In the gain term we substitute $E' = E/(1-y)$, which turns $E^s$ into $(1-y)^s E'^s$, and the moments $M_s(\ell) = \int \dd E \, E^s \phi(\ell, E) = \langle E^s \rangle$ obey
\begin{equation}
  \frac{\dd M_s}{\dd \ell} = -\Phi(s) \, M_s
  \label{eq:momenteq}
\end{equation}
for any starting distribution. The delta function gives $M_s(0) = E_0^s$, and Eq.~(\ref{eq:momentlaw}) follows. Two values of $s$ are familiar. At $s = 0$ the equation says that collisions degrade muons without removing any, since $\Phi(0) = 0$, and at $s = 1$ it is Eq.~(\ref{eq:dedx}). Note, that we will never need $\phi$ itself. When we need it, for the loss distributions in Table~\ref{tab:losslaw}, we recover it by inverting Eq.~(\ref{eq:genfn}) numerically (Appendix~\ref{app:diag}).

The same substitution works on a spectrum. Applied to a power law $E^{-1-s}$, the operator returns the same power law
\begin{equation}
    \Cqed\!\left[ E^{-1-s} \right] = - \Phi(s) \, E^{-1-s},
    \label{eq:eigen}
\end{equation}
so a spectrum of index $s$ propagates over a column depth $\ell$ by a multiplication by $e^{-\ell\Phi(s)}$. Cascade theory and cosmic-ray physics use Eq.~(\ref{eq:Phi}) at a fixed index as this attenuation rate, the Z~moment~\cite{Rossi:1941zza, Lipari:1993hd, Gaisser:2016uoy}.

The muon sources measured by neutrino telescopes are such a power law. A flux $\phi_\nu \propto E_\nu^{-\gamma}$ and a cross section $\sigma_{\nu N} \propto E_\nu^{\lambda}$~\cite{Gandhi:1998ri, Connolly:2011vc, Cooper-Sarkar:2011jtt, Bertone:2018dse} make a muon source of effective index
\begin{equation}
    A \equiv \gamma - \lambda - 1,
    \label{eq:Adef}
\end{equation}
because the weak vertex rescales the energy the way a collision does (Appendix~\ref{app:diag}). Both indices run slowly, so we read $A$ as the local logarithmic slope of $\phi_\nu \sigma_{\nu N}$. One function therefore describes both the stochastic loss of a single muon, through Eq.~(\ref{eq:genfn}) at a chosen $s$, and the attenuation of a power-law spectrum, through Eq.~(\ref{eq:eigen}) at its index $A$. Appendix~\ref{app:diag} solves the transport equation directly and shows why the two share one function.

\subsection{What We Need to Know about the Collisions}
\label{sec:collisions}

To evaluate $\Phi$ we need the loss spectrum $\dd\Gamma/\dd y$, but each quantity in this paper needs it only through a few of its integrals. Any spectrum with the same integrals gives the same answer for that quantity, so we never need a unique reconstruction of the physical loss spectrum. The mean energy needs $b_\mu$. The muon range needs the two logarithmic moments of Eq.~(\ref{eq:Phimoments}). Finally, $\Phi$ at a positive integer $n$ needs only the first $n$ moments of $y$, because the binomial in Eq.~(\ref{eq:Phi}) then terminates
\begin{eqnarray}
    \Phi(1) &=& b_\mu, \label{eq:exact1} \\
    \Phi(2) &=& 2 b_\mu - d_\mu, \label{eq:exact2} \\
    \Phi(3) &=& 3 b_\mu - 3 d_\mu + t_\mu, \label{eq:exact3}
\end{eqnarray}
where $d_\mu = \langle y^2 \rangle$ and $t_\mu = \langle y^3 \rangle$ are the second and third moments per unit column depth, defined like $b_\mu = \langle y \rangle$ in Eq.~(\ref{eq:bdef}). Three moments therefore fix $\Phi(1)$, $\Phi(2)$ and $\Phi(3)$ for every spectrum that shares them, and we take those moments from PROPOSAL~\cite{Koehne:2013gpa, Dunsch:2018nsc}.

\begin{figure}[t]
\centering
\includegraphics[width=\columnwidth]{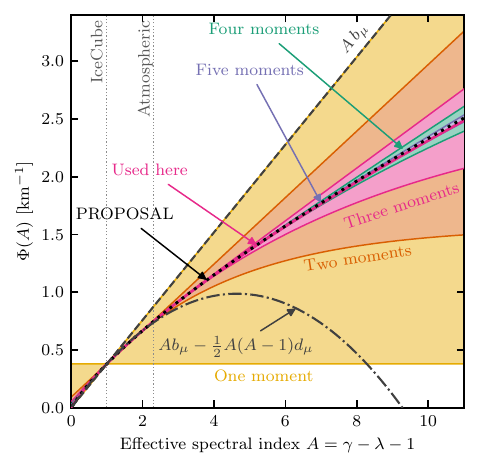}
\caption{$\Phi$ as a function of the effective spectral index. Each colored band is the full range of $\Phi(A)$ over every positive loss spectrum with the first $n$ moments of PROPOSAL's spectrum, whatever its shape, and it pinches at the integers $1 \le A \le n$. The solid pink line is the family of Eq.~(\ref{eq:betafam}), the dotted line is PROPOSAL, the dashed line is the first-order term $A b_\mu$, and the dash-dotted line is the drift-diffusion form of Refs.~\cite{Palmisano:2025abd, Palmisano:2026sid}.}
\label{fig:exponent}
\end{figure}

Figure~\ref{fig:exponent} shows how little the moments leave free between the integers. Its bands are the full range of $\Phi(A)$ over every positive spectrum with a given number of moments, whatever its shape. With three moments the band pinches to a line at $A = 1$, $2$ and $3$, and between them it stays within half a percent. It opens to $+5\%$ at $A = 0.5$, and to $-15\%$ and $+9\%$ at $A = 10$. That width measures what three moments leave undetermined. It says nothing about whether a particular spectrum is wrong.

The moments of $y$ do not fix $\Phi'(0)$ and $\Phi''(0)$ at all. % A spectrum can carry the same three moments with part of its rate at $y = 1$, a collision that takes all of the energy, and there $-\ln(1-y)$ is infinite. The moments of $y$ cannot tell a collision that takes $99.99\%$ of the energy from one that takes all of it, while the logarithm can. 
For the muon range, we therefore take $\Phi'(0)$ and $\Phi''(0)$ from PROPOSAL's tabulated spectrum.

Where we need $\Phi$ as a whole function, for Figure~\ref{fig:exponent}, Table~\ref{tab:losslaw} and the energy of a single event in Section~\ref{sec:energy}, we use a convenient spectrum with PROPOSAL's first three moments. Any such spectrum would do as well at the integers, and we choose ours with the physics of the cross sections in mind.

Radiative spectra are close to power laws in $y$. Bremsstrahlung falls as $1/y$, and pair production, which dominates at small $y$, falls more steeply, as $y^{-1.7}$ to $y^{-2}$ over the decades of $y$ that carry most of $b_\mu$. Near $y = 1$ bremsstrahlung and photonuclear scattering stay finite up to the kinematic end point. The family
\begin{equation}
  \frac{\dd\Gamma}{\dd y} = \kappa\, y^{q-1}(1-y)^p
  \label{eq:betafam}
\end{equation}
has both limits built in, a power law at small $y$ and a power of $(1-y)$ at the hard end. Matched to PROPOSAL's first three moments at $1$~PeV it returns $q - 1 = -1.74$, the slope of the spectrum where $b_\mu$ is made, and $p = -0.18$, a hard end that stays nearly flat up to $y = 1$. For this family $\Phi$ is a difference of Beta functions
\begin{equation}
  \Phi(s) = \kappa \left[ B(q, p+1) - B(q, p+1+s) \right].
  \label{eq:phithree}
\end{equation}
It tracks PROPOSAL's $\Phi$ to about half a percent across the energies and spectral indices of interest. At large $A$, where the moments no longer fix $\Phi$, it still follows PROPOSAL, because $\Phi$ there grows as $A^{-q} = A^{0.74}$ and is set by the soft end the family shares with the cross sections.

%%%%%%%%%%%%%%%%%%%%%%%%%%%%%%%%%%%%%%%%%%%%%%%%%%%%%%
%%%%%%%%%%%%%%%%%%%%%%%%%%%%%%%%%%%%%%%%%%%%%%%%%%%%%%

\subsection{Relation to the Drift-Diffusion Expansion}
\label{sec:comparison}

Refs.~\cite{Palmisano:2025abd, Palmisano:2026sid} use the same moments differently. They expand the collision operator in $y$ and truncate the expansion, which turns the moments of the loss spectrum into the coefficients of an approximate operator. Here, we instead use the moments as a constraint on arbitrary functions. So any function (loss spectrum) that meets those constraints gives the same answer for the observables we want. For $\Phi$ their~\cite{Palmisano:2025abd, Palmisano:2026sid} truncation reads
\begin{equation}
    \Phi(A) = A b_\mu - \frac{A(A-1)}{2} d_\mu + \dots.
    \label{eq:km}
\end{equation}

The truncated $\Phi$ describes a logarithmic loss that can decrease, a muon that gains energy (Appendix~\ref{app:fp}), which no positive loss spectrum allows. Its curve therefore lies outside the two-moment band of Figure~\ref{fig:exponent} everywhere except at $A = 0$, $1$ and $2$, and it turns negative above $A = 9.3$ in water at $1$~PeV.

Where the series terminates the truncation is right. It reproduces Eqs.~(\ref{eq:exact1}) and (\ref{eq:exact2}), and the IceCube diffuse index and cross-section slope give $A = 0.98$ near $10$~PeV~\cite{IceCube:2016umi, Abbasi:2021qfz, Bertone:2018dse}, almost on the first of them. The drift-diffusion form therefore works well for the astrophysical flux measured by IceCube. What it misses, is the tail of the loss distribution, which Section~\ref{sec:validation} tests, and the spread of the muon range, because it carries $d_\mu = \langle y^2 \rangle$ where the muon range needs $\langle \ln^2(1-y) \rangle$.

\subsection{The Muon Range}
\label{sec:range}

With the statistics of $W$ in hand, we can now say how far a muon travels. The muon crosses the threshold when $W$ first exceeds $w = \ln(E_0/E_{\mathrm{thr}})$, and the mean column depth at which that happens is the muon range
\begin{equation}
  L(E_0) = \frac{\ln (E_0/E_\mathrm{thr})}{\Phi'(0)} - \frac{\Phi''(0)}{2 \, \Phi'(0)^2},
  \label{eq:Lclosed}
\end{equation}
which Appendix~\ref{app:renewal} derives together with its variance. Compared to Eq.~(\ref{eq:range_standard}) two things have changed. First, the coefficient has moved from $b_\mu = \Phi(1)$ to $\Phi'(0)$, from the rate of the mean energy to the rate of its logarithm. This shortens the leading term by $b_\mu / \Phi'(0) = 0.78$ in water at $1$~PeV, because a hard collision costs far more in the logarithm than in the fraction. Second, a constant has been added. The muon jumps across the threshold, and a large step is more likely to straddle a level than a small one, so on average it lands a factor of $1.4$ in energy below the threshold, its mean overshoot. The constant is the column depth the muon spends waiting for that collision, $0.8$~km of water at any threshold, and it adds a little to the muon range where the coefficient removes more. We take $\Phi'(0)$ and $\Phi''(0)$ from the quadrature of Section~\ref{sec:collisions}.

\subsection{Validating the Transport}
\label{sec:validation}

We test the transport against PROPOSAL~\cite{Koehne:2013gpa, Dunsch:2018nsc, Garcia:2020jwr, Garg:2022ugd}, for muons of $1$~PeV in water unless stated. The mean energy loss over a kilometer is no test. Our transport and the drift-diffusion one both reproduce it to $0.5\%$, because the mean is the identity $\Phi(1) = b_\mu$ that every treatment carries by construction. The energy-loss distribution is the benchmark~\cite{Lipari:1991ut, Dutta:2000hh, Safa:2021ghs}. Once the muon has lost a factor of $4.5$ in energy, the second-order expansion is low by a factor of $330$, and by fifteen orders of magnitude at a factor of $20$, while numerically inverting Eq.~(\ref{eq:genfn}) tracks the Monte Carlo throughout (Table~\ref{tab:losslaw}).

We also test the muon range directly, by propagating muons in PROPOSAL until they cross a $10$~TeV threshold. The sampled muon ranges match Eq.~(\ref{eq:Lclosed}) to within $3\%$ at every production energy from $10^5$ to $10^8$~GeV. The continuous-slowing-down muon range runs $10$ to $25\%$ too long over the same energies, and the drift-diffusion form returns the spread of the muon range $19$ to $40\%$ too low (Appendix~\ref{app:renewal}).

The muon range of Eq.~(\ref{eq:Lclosed}) depends on the medium and the loss spectrum alone, and nothing about a detector enters it. Section~\ref{sec:volume} turns it into an effective area that a reader can evaluate from a detector's geometry and two instrument numbers.

%%%%%%%%%%%%%%%%%%%%%%%%%%%%%%%%%%%%%%%%%%%%%%%%%%%%%%%%%%%%%%%%%%%%%%%%%%%%%%%%%%%%%%%%%%%
%%%%%%%%%%%%%%%%%%%%%%%%%%%%%%%%%%%%%%%%%%%%%%%%%%%%%%%%%%%%%%%%%%%%%%%%%%%%%%%%%%%%%%%%%%%

\section{The Effective Area}
\label{sec:volume}

Having derived the muon range, we can now calculate the effective volume of a neutrino telescope. A muon counts if it is born inside the detector or within one muon range of it, so the effective volume is
\begin{equation}
  V_{\mathrm{eff}} = A_{\mathrm{proj}} \, L(E_0) + V_{\mathrm{det}},
  \label{eq:volume}
\end{equation}
where $A_{\mathrm{proj}}$ is the projected area of the instrument, $L$ the muon range of Eq.~(\ref{eq:Lclosed}), and $V_{\mathrm{det}}$ the instrumented volume. Equation~(\ref{eq:volume}) is the standard effective volume~\cite{Gaisser:2016uoy, Palmisano:2025abd, Palmisano:2026sid}, and the transport supplies $L$. We evaluate it at $E_0 = (1 - \langle y_w \rangle) E_\nu$, the energy a charged-current interaction leaves the muon, with $\langle y_w \rangle = 0.20$~\cite{Gandhi:1998ri, IceCube:2018pgc}.

\subsection{The Estimate}
\label{sec:closedform}
\label{sec:response}

A published effective area is an average over its solid angle, so Eq.~(\ref{eq:estimate}) with the volume of Eq.~(\ref{eq:volume}) becomes
\begin{eqnarray}
  \langle A_{\mathrm{eff}} \rangle(E_\nu) &\simeq& \epsilon_0 \left( \frac{E_\nu}{1~\mathrm{PeV}} \right)^{k} n_N \, \sigma_{\mathrm{CC}}(E_\nu) \, T(E_\nu) \nonumber \\
  && \times \left[ \langle A_{\mathrm{proj}} \rangle \, L(E_\nu) + V_{\mathrm{det}} \right], \label{eq:skyavg} \\
  L(E_\nu) &=& 2.17~\mathrm{km} \, \ln\!\left( \frac{0.8 \, E_\nu}{E_{\mathrm{thr}}} \right) + 0.81~\mathrm{km}, \nonumber
\end{eqnarray}
with $L$ set to zero below threshold. Here $\epsilon_0$ is the efficiency $\epsilon$ of Eq.~(\ref{eq:estimate}) at its high-energy plateau, and $T = \langle e^{-X/\Lambda_\nu} \rangle_\Omega$ is the averaged probability that the neutrino survives the column depth $X$ to the detector, with $\Lambda_\nu = 1/(n_N \sigma_{\mathrm{tot}})$ its interaction length and $\sigma_{\mathrm{tot}}$ the total, charged- plus neutral-current, cross section~\cite{IceCube:2017roe, Vincent:2017svp, Dziewonski:1981xy}. The mean projected area $\langle A_{\mathrm{proj}} \rangle$ is a quarter of the instrument's surface area. The power $k$ accounts for the halo, the muons that pass outside the instrumented volume but close enough to trigger it, and Section~\ref{sec:twonumbers} relates it to the reach.

The two lengths in $L$, both in kilometers of water equivalent, are where the transport enters. The leading one is $1/\Phi'(0)$, the rate at which the logarithm of the energy falls, where the standard muon range of Eq.~(\ref{eq:range_standard}) carries $1/b_\mu$, and the second is the mean overshoot of Section~\ref{sec:range}. We take both from Table~\ref{tab:renewal} at $100$~TeV, which holds the muon range within $6\%$ of the one that follows the loss spectrum down the descent (Appendix~\ref{app:renewal}).

At a sea site a downgoing muon can only be born in the water above the detector. We therefore average the two hemispheres separately, each with its own $T$, and cap the downgoing muon range at $d/\cos\theta$ for a detector at depth $d$, whose hemisphere mean is $d\,[1 + \ln(L/d)]$ once $L$ exceeds $d$.

\subsection{Two Detector Parameters}
\label{sec:twonumbers}

Besides its geometry and the normalization $\epsilon_0$, held at the collaboration's value, two numbers describe a detector. The first is the threshold $E_{\mathrm{thr}}$, one effective energy standing for a selection that turns on over a range of them. The second is the reach, $\Lambda$. A bright muon triggers the detector from farther away than a faint one, so the instrument's radius grows by $2.3\,\Lambda$ per order of magnitude in muon energy around $E_{\mathrm{piv}} = 1$~PeV. Cherenkov light scales with the energy a muon sheds~\cite{Groom:2001kq, IceCube:2013dkx}, so $\Lambda$ is the attenuation length of the site's water or ice, predicted from its measured optics in Appendix~\ref{app:response}. A quality selection removes the halo tracks, and $\Lambda$ vanishes at analysis level. To first order in $\Lambda$ the halo is the power of Eq.~(\ref{eq:skyavg})
\begin{equation}
  k = \Lambda \, \frac{R_{\mathrm{det}} + c\,h/4}{R_{\mathrm{det}}^2/2 + c\,R_{\mathrm{det}}\,h/4},
  \label{eq:kreach}
\end{equation}
for an instrument of radius $R_{\mathrm{det}}$, height $h$ and side coefficient $c$, which is $2$ for a cylinder (Appendix~\ref{app:recipe}). The power reproduces the full growth of the projected area, Eq.~(\ref{eq:reach}) below, to $7\%$, and $\langle A_{\mathrm{proj}} \rangle$ is then the footprint alone. Every other input is geometry and the transmission, tabulated in Appendix~\ref{app:inputs} with $n_N = 6.1 \times 10^{23}$~cm$^{-3}$ for water and BGR18 for $\sigma_{\mathrm{CC}}$~\cite{Bertone:2018dse}.

\begin{figure}[t]
\centering
\includegraphics[width=\columnwidth]{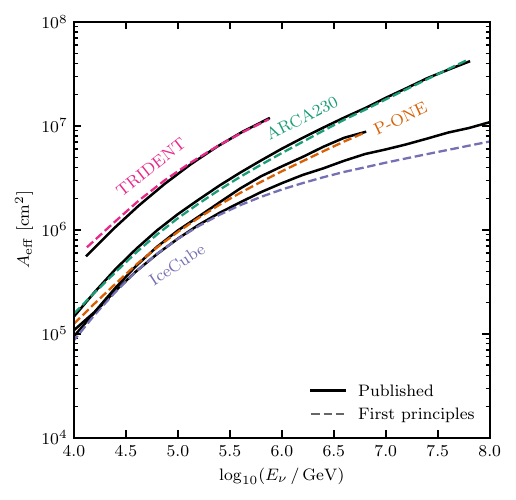}
\caption{Four published effective areas: the IceCube DR2 upgoing sample~\cite{Abbasi:2026ehs}, the KM3NeT/ARCA230 trigger-level curve~\cite{KM3NeT:2024paj}, the effective area published with the P-ONE proposal~\cite{P-ONE:2020ljt}, and TRIDENT's simulated track response~\cite{TRIDENT:2022hql, Morton-Blake:2025vpp} (black) compared to the estimate of Eq.~(\ref{eq:skyavg}).}
\label{fig:skyavg}
\end{figure}

Figure~\ref{fig:skyavg} evaluates Eq.~(\ref{eq:skyavg}) with nothing fitted. The normalization is the collaboration's, the reach is what each site's optics predict (Appendix~\ref{app:response}), and the threshold is a $300$~GeV trigger at the sea sites. ARCA230, the full KM3NeT/ARCA detector of $230$ detection units, and P-ONE come out flat and $7$ to $8\%$ below the published curves, and that shortfall is the physics the estimate leaves out and Section~\ref{sec:theresponse} restores.

The inputs the geometry does not supply are properties of the instruments. IceCube's table is at analysis level, with its turn-on at $3.5$~TeV and no reach, and the estimate falls to $0.75$ of it at $10$~PeV. P-ONE's clusters are $120$~m across, too compact for the bulk optics to set the halo, so its reach is ARCA's relative to the radius, a tenth of it. TRIDENT's map fixes neither its selection level nor its normalization, and $\epsilon_0 = 0.7$ is the value its table returns. Without the muon range, Eq.~(\ref{eq:estimate}) with the instrumented volume falls short of every table by a factor of $6$ to $47$ and misses the slope everywhere.

%%%%%%%%%%%%%%%%%%%%%%%%%%%%%%%%%%%%%%%%%%%%%%%%%%%%%%
%%%%%%%%%%%%%%%%%%%%%%%%%%%%%%%%%%%%%%%%%%%%%%%%%%%%%%

%%%%%%%%%%%%%%%%%%%%%%%%%%%%%%%%%%%%%%%%%%%%%%%%%%%%%%%%%%%%%%%%%%%%%%%%%%%%%%%%%%%%%%%%%%%
%%%%%%%%%%%%%%%%%%%%%%%%%%%%%%%%%%%%%%%%%%%%%%%%%%%%%%%%%%%%%%%%%%%%%%%%%%%%%%%%%%%%%%%%%%%

\section{Improving the Estimate}
\label{sec:theresponse}
\label{sec:tau}

Equation~(\ref{eq:skyavg}) makes three approximations, and we give what each is worth at KM3NeT/ARCA230. It propagates every muon through water, where an upgoing muon is born in rock, which sheds energy faster per unit column depth and shortens the muon range, a move of under $2\%$ (Appendix~\ref{app:renewal}). It treats the Earth as an absorber, where a neutrino that scatters neutral-current is not lost~\cite{Gandhi:1998ri, Vincent:2017svp}. In reality, it continues at lower energies, contributing $4$ to $9\%$ (Appendix~\ref{app:regeneration}). It also doesn't include tau neutrinos, which produce a tau that can decay to a muon~\cite{Halzen:1998be, Dutta:2000hh, Bugaev:2003sw, Alvarez-Muniz:2018owm, Safa:2021ghs}, worth $13$ to $18\%$ and flat in energy, see Appendix~\ref{app:tau}.

Including these effects, and the arrival direction, the response is
\begin{eqnarray}
  A_{\mathrm{eff}}(E_\nu, \Omega) &=& \epsilon_0 \, n_N \sum_k w_k(E_\nu, \Omega) \, \sigma_{\mathrm{CC}}(E_k) \nonumber \\
  && \times \left[ A_{\mathrm{proj}}(\Omega, E_k) \, L(E_k) + V_{\mathrm{det}} \right],
  \label{eq:aeff}
\end{eqnarray}
where the sum runs over the energies at which the neutrino can arrive. A neutrino that scatters neutral-current continues at a lower energy, so it arrives on a ladder $E_k = E_\nu (1 - \langle y \rangle_{\mathrm{NC}})^k$, with $\langle y \rangle_{\mathrm{NC}} = 0.25$ the mean neutral-current inelasticity, and $w_k(E_\nu, \Omega)$ is the probability that it reaches the detector on rung $k$ (Appendix~\ref{app:regeneration}). The muon is born at $E_0 = (1 - \langle y_w \rangle) E_k$, $L$ is the two-medium muon range of Eq.~(\ref{eq:Ltwomedium}), and the tau source adds a sum of its own. $A_{\mathrm{proj}}(\Omega, E)$ is the projected area of Appendix~\ref{app:recipe} at the reach radius
\begin{equation}
  R_{\mathrm{eff}}(E) = \max\!\left[ 0, \quad R_{\mathrm{det}} + \Lambda \ln\!\left( E / E_{\mathrm{piv}} \right) \right],
  \label{eq:reach}
\end{equation}
whose maximum lets the reach turn negative below $E_{\mathrm{piv}}$ and so covers the turn-on as well. Equation~(\ref{eq:skyavg}) is Eq.~(\ref{eq:aeff}) with one rung, one source, the sky average taken, and the halo expanded to first order.

%%%%%%%%%%%%%%%%%%%%%%%%%%%%%%%%%%%%%%%%%%%%%%%%%%%%%%%%%%%%%%%%%%%%%%%%%%%%%%%%%%%%%%%%%%%
%%%%%%%%%%%%%%%%%%%%%%%%%%%%%%%%%%%%%%%%%%%%%%%%%%%%%%%%%%%%%%%%%%%%%%%%%%%%%%%%%%%%%%%%%%%

\section{Does It Work?}
\label{sec:aeff}

\subsection{Published Effective Areas}

We now compare the full model with four published effective areas: the IceCube DR2 upgoing table~\cite{Abbasi:2026ehs}, the KM3NeT/ARCA230 trigger-level curve~\cite{KM3NeT:2024paj}, the effective area published with the P-ONE proposal~\cite{P-ONE:2020ljt}, and TRIDENT's simulated track response~\cite{TRIDENT:2022hql, Morton-Blake:2025vpp}. We evaluate Eq.~(\ref{eq:aeff}) at every tabulated energy and direction, per unit parent neutrino energy. The physics is fixed throughout: the BGR18 cross section~\cite{Bertone:2018dse}, $\langle y_w \rangle = 0.20$~\cite{Gandhi:1998ri, IceCube:2018pgc}, PROPOSAL's loss spectrum, and the instrument geometry of Appendix~\ref{app:inputs}. We hold the selection normalization at the collaboration's level, unity for the trigger-level tables and the analysis-level plateau for DR2 (Appendix~\ref{app:response}), and it is free for TRIDENT alone. The threshold and the reach are fitted to each table under an assumed $5\%$ error per tabulated point.

%%%%%%%%%%%%%%%%%%%%%%%%%%%%%%%%%%%%%%%%%%%%%%%%%%%%%%
\begin{figure}[t]
\centering
\includegraphics[width=\columnwidth]{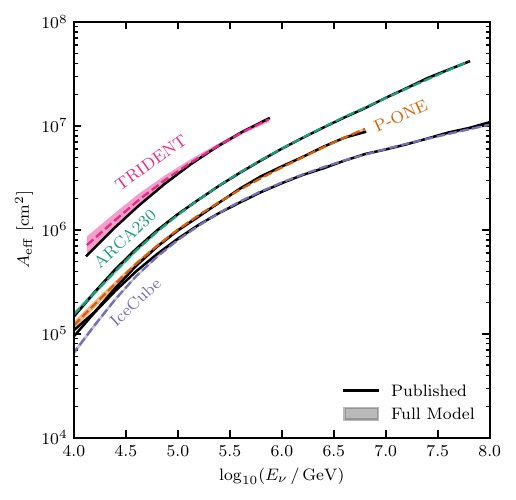}
\caption{Four published effective areas: the IceCube DR2 upgoing sample~\cite{Abbasi:2026ehs}, the KM3NeT/ARCA230 trigger-level curve~\cite{KM3NeT:2024paj}, the effective area published with the P-ONE proposal~\cite{P-ONE:2020ljt}, and TRIDENT's simulated track response~\cite{TRIDENT:2022hql, Morton-Blake:2025vpp} (black) compared to the full model of Eq.~(\ref{eq:aeff}).}
\label{fig:aeff}
\end{figure}
%%%%%%%%%%%%%%%%%%%%%%%%%%%%%%%%%%%%%%%%%%%%%%%%%%%%%%

Figure~\ref{fig:aeff} shows the four comparisons. Inside each fit window the published curve and the posterior median agree in level to within half a percent at three sites and to $4.5\%$ at TRIDENT, with percent-level scatter about that level. The two numbers behind them are Table~\ref{tab:recipe}. The threshold lands on each selection's turn-on, $250$ to $310$~GeV at trigger level and $3.5$~TeV for IceCube's analysis-level table, and the reach lands on the optics prediction at the three trigger-level sea sites (Appendix~\ref{app:response}), a factor of two below it at P-ONE, whose clusters are too compact for the bulk optics. The transport's own uncertainty is photonuclear scattering, $\pm 28\%$ on the mean loss at $100$~PeV, and after propagation through the muon range it moves an effective area by a few percent (Appendix~\ref{app:response}). Every band in the figures below is this envelope.

% Fitting the estimate of Eq.~(\ref{eq:skyavg}) instead moves the missing physics into the instrument numbers, a $37$~m reach at IceCube's analysis-level table and thresholds at the prior floor at the sea sites. The kernel does not enter. A rescaled kernel is absorbed by the threshold and the normalization, so no published table constrains it independently of the detector. 

%%%%%%%%%%%%%%%%%%%%%%%%%%%%%%%%%%%%%%%%%%%%%%%%%%%%%%
\begin{table*}[t]
\caption{The instrument numbers of the four sites, fitted by the estimate of Eq.~(\ref{eq:skyavg}) and by the full model of Eq.~(\ref{eq:aeff}), with the physics fixed at PROPOSAL's loss spectrum and the BGR18 cross section. The selection normalization is fixed at the collaboration's level except at TRIDENT.}
\begin{center}
% Machine-written by docs/make_recipe_table.py -- do not edit by hand.
% Sources: scripts/2026_muon_transport/output/77_chains_sigma05.npz (IceCube, ARCA230, P-ONE; physics fixed, eps_0 held) and 82_trident2025_chain.npz (TRIDENT 2025 map, eps_0 free).
\begin{tabular}{lccccc}
\hline\hline
detector & $\varepsilon_0$ & $E_{\mathrm{thr}}$ [GeV] & $\Lambda$ fitted [m] & $\Lambda$ predicted [m] & $A_{\mathrm{eff}}(1~\mathrm{PeV})$ [m$^2$] \\
\hline
IceCube & $0.956$ (fixed) & $3550^{+228}_{-222}$ & $7 \pm 4$ & $30$ to $59$ & $285$ \\
ARCA230 & $1$ (fixed) & $313^{+53}_{-47}$ & $45 \pm 3$ & $41$ to $50$ & $604$ \\
P-ONE & $1$ (fixed) & $250^{+48}_{-41}$ & $11 \pm 1$ & $24$ to $30$ & $398$ \\
TRIDENT & $0.72^{+0.08}_{-0.07}$ (free) & $248^{+203}_{-98}$ & $38^{+59}_{-57}$ & $15$ to $27$ & $1170$ \\
\hline\hline
\end{tabular}

\end{center}
\label{tab:recipe}
\end{table*}
%%%%%%%%%%%%%%%%%%%%%%%%%%%%%%%%%%%%%%%%%%%%%%%%%%%%%%

\subsection{The Event Sample}
\label{sec:events}

Nothing in this subsection is fitted to the events. We fold the responses through the smearing matrices released for IC86, the full $86$-string IceCube configuration, into reconstructed muon energy. Every flux is fixed from outside: the MCEq conventional and prompt atmospheric fluxes at their nominal normalizations~\cite{Fedynitch:2015zma}, and IceCube's nine-and-a-half-year tracks measurement~\cite{Abbasi:2021qfz} at an equal flavor ratio through both track channels. The tracks flux was itself measured on an overlapping event sample, so the benchmark tests the response at a given flux, not the flux.

%%%%%%%%%%%%%%%%%%%%%%%%%%%%%%%%%%%%%%%%%%%%%%%%%%%%%%
\begin{figure}[t]
\centering
\includegraphics[width=\columnwidth]{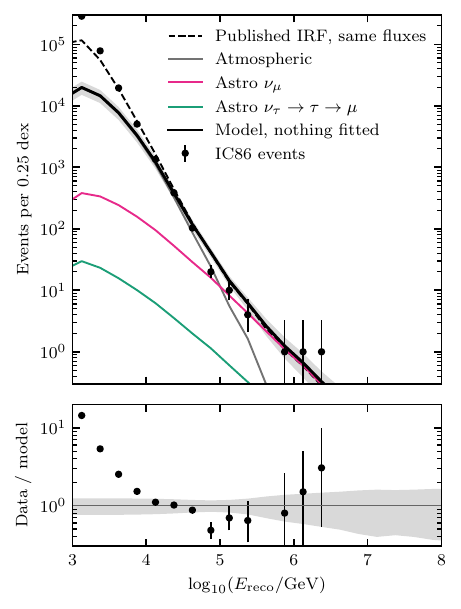}
\caption{Reconstructed-energy spectrum of the upgoing IC86 sample as measured (points) and predicted by our model (lines). The dashed line is the same prediction using IceCube's released effective area.}
\label{fig:events}
\end{figure}
%%%%%%%%%%%%%%%%%%%%%%%%%%%%%%%%%%%%%%%%%%%%%%%%%%%%%%

Figure~\ref{fig:events} shows the prediction against the $10.7$-year upgoing IC86 sample of DR2. Above the turn-on region of the response the model predicts $565 \pm 109$ events against $526$ observed, three quarters of them atmospheric, with the band set by the uncertainty on the conventional normalization. The same prediction through IceCube's own released effective area gives $629$, and the two responses track each other across three decades of reconstructed energy. Where the data fall below both, the deficit is shared by the two responses and belongs to the fluxes. The declination distribution is tracked across all ten bands (Appendix~\ref{app:events}). The model also predicts the flavor composition, and at an equal flavor ratio $6.5\%$ of the astrophysical tracks in the window come from $\nu_\tau \to \tau \to \mu$, compared with the ${\sim}5\%$ IceCube's own track fits assume for the same channel~\cite{IceCube:2016umi, Abbasi:2021qfz}.

%%%%%%%%%%%%%%%%%%%%%%%%%%%%%%%%%%%%%%%%%%%%%%%%%%%%%%%%%%%%%%%%%%%%%%%%%%%%%%%%%%%%%%%%%%%
%%%%%%%%%%%%%%%%%%%%%%%%%%%%%%%%%%%%%%%%%%%%%%%%%%%%%%%%%%%%%%%%%%%%%%%%%%%%%%%%%%%%%%%%%%%

\section{Three Uses of the Response}
\label{sec:applications}

The full model of Eq.~(\ref{eq:aeff}) runs per arrival direction in a fraction of a second, so it can be pointed where a simulation per point is costly.

\subsection{Point Sources at Four Sites}
\label{sec:pointsource}

A point source fixes the declination, and the full model follows it per direction around the daily rotation of the sky~\cite{IceCube:2019cia, IceCube:2022der}. For $\phi_\nu(E) = \phi_0 (E/E_{\mathrm{piv}})^{-\gamma}$ the expected count is linear in $\phi_0$, so the limit is a single integral
\begin{equation}
  \phi_0^{\mathrm{lim}}(\delta) = \frac{N_{\mathrm{lim}}}
  {T \displaystyle\int_{E_{\min}} \dd E \quad A_{\mathrm{eff}}(E, \delta)
   \left( E / E_{\mathrm{piv}} \right)^{-\gamma}},
  \label{eq:pslim}
\end{equation}
where $N_{\mathrm{lim}}$ is the Feldman-Cousins $90\%$ upper limit averaged over background-only outcomes~\cite{Feldman:1997qc} and $T$ the exposure time. We count the background from an MCEq atmospheric flux through the same response, inside the bin IceCube's released point spread gives at each energy and direction. We choose $E_{\min}$ at each declination to minimize the limit. Against IceCube's fourteen-year track sensitivity the result agrees to about $20\%$ across the northern sky (Appendix~\ref{app:pointsource}).

\begin{figure}[t]
\centering
\includegraphics[width=\columnwidth]{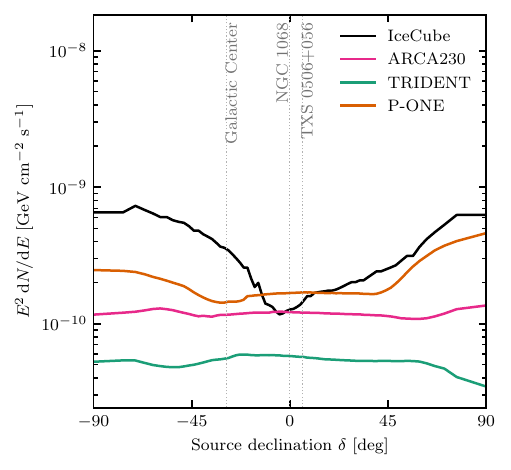}
\caption{Point-source limit of Eq.~(\ref{eq:pslim}) against source declination for four sites~\cite{IceCube:2016zyt, KM3NeT:2018wnd, P-ONE:2020ljt, TRIDENT:2022hql}, on a common fourteen-year exposure and $\gamma = 2$, with the atmospheric background counted in the bin the point spread gives and the window chosen at each declination.}
\label{fig:ps_sites}
\end{figure}

Figure~\ref{fig:ps_sites} shows the same limit at four sites on a common fourteen-year exposure, computed in one pass with a common threshold and one fitted reach per site. IceCube's limit varies by a factor of $6.2$ across the sky against a factor of $1.3$ at ARCA230. A polar site holds every source at a fixed zenith, while an equatorial site sweeps each through both regimes daily.

%%%%%%%%%%%%%%%%%%%%%%%%%%%%%%%%%%%%%%%%%%%%%%%%%%%%%%
%%%%%%%%%%%%%%%%%%%%%%%%%%%%%%%%%%%%%%%%%%%%%%%%%%%%%%

\subsection{The Energy of One Event}
\label{sec:energy}

What is the neutrino energy behind a single track? For a muon born anywhere along a uniform column, the column depth it spends in each interval of logarithmic loss is a fixed density, $u(w)$. Its transform is $1/\Phi(s)$ (Appendix~\ref{app:renewal}), and here we need $\Phi$ as a whole function, so we take it from the family of Eq.~(\ref{eq:betafam}). We fold $u$ with the reported muon energy of $120^{+110}_{-60}$~PeV~\cite{KM3NeT:2025npi}, the survival along the sea path, and a flux prior.

\begin{figure}[t]
\centering
\includegraphics[width=\columnwidth]{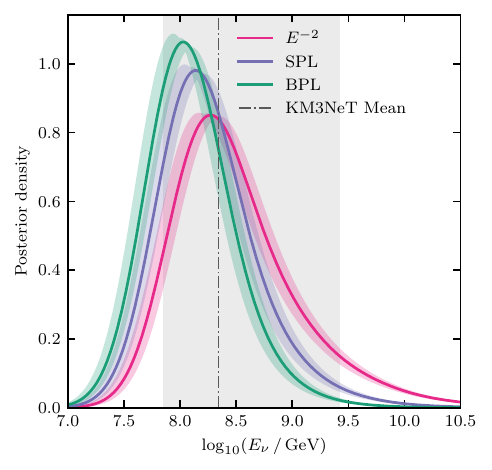}
\caption{Neutrino-energy posterior for KM3-230213A from the potential density of the loss spectrum of Eq.~(\ref{eq:betafam}), matched to PROPOSAL, under three flux priors, each with the loss-model envelope drawn as a band. KM3NeT's published $E^{-2}$ median and $90\%$ interval~\cite{KM3NeT:2025npi} are marked for comparison.}
\label{fig:event_energy}
\end{figure}

Figure~\ref{fig:event_energy} shows the neutrino-energy posterior under three choices of that prior, each with the loss-model envelope as a band. Here, the published broken power law (BPL)~\cite{Naab:2023xcz} and single power law (SPL)~\cite{Abbasi:2021qfz} lines are IceCube fits. Under KM3NeT's own $E^{-2}$ prior the median is $250$~PeV, beside the collaboration's $220$~PeV~\cite{KM3NeT:2025npi}, and both $90\%$ intervals span two decades. The flux prior moves the median more than the loss spectrum does.

%%%%%%%%%%%%%%%%%%%%%%%%%%%%%%%%%%%%%%%%%%%%%%%%%%%%%%
%%%%%%%%%%%%%%%%%%%%%%%%%%%%%%%%%%%%%%%%%%%%%%%%%%%%%%

\subsection{Consistency between Detectors}
\label{sec:tension}

Is the event consistent with the IceCube diffuse flux? With the full model at both experiments, the question becomes a likelihood comparison on the two flux parameters, which separates what the detectors contribute from what the event does. The IceCube side convolves the DR2 upgoing table with that release's $13.6$~years of livetime~\cite{Abbasi:2026ehs}. The ARCA21 side, the $21$-unit configuration that recorded the event, convolves the released bright-track area~\cite{KM3NeT:2025ccp, Li:2025tqf} with the full ARCA21 livetime. Note, this is not the effective area shown in Figure~\ref{fig:aeff}, which was for ARCA230 and at trigger level. The event enters through its energy likelihood and a second Poisson factor asserting that no other bright track appeared between $10$~PeV and $10$~EeV. We use likelihood ratios for both regions, since with one event a posterior would pile up against whatever prior range is imposed.

\begin{figure}[t]
\centering
\includegraphics[width=\columnwidth]{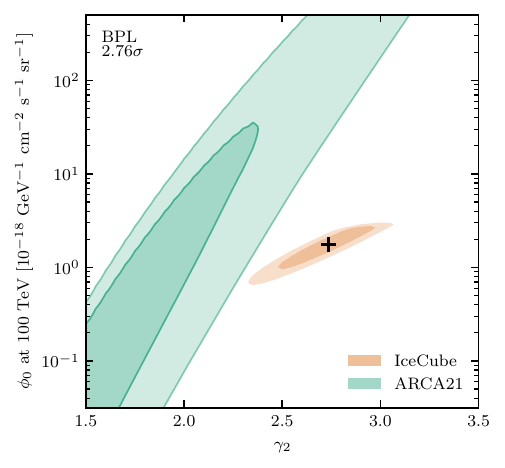}
\caption{Constraints on a diffuse flux from the IceCube through-going sample and from KM3-230213A under IceCube's published broken power law, as $68\%$ (darker) and $95\%$ (lighter) two-parameter likelihood ratios. The IceCube region is a forecast at the median expected counts, and the ARCA21 region combines the energy likelihood with the loss-model ensemble.}
\label{fig:flux_tension}
\end{figure}

Figure~\ref{fig:flux_tension} shows the two constraints on the same flux parameters, under IceCube's published broken power law~\cite{Naab:2023xcz}, where the datasets are in tension at $2.8\sigma$. Under a single power law at the diffuse fit the tension is $2.24\sigma$ (Appendix~\ref{app:tension}).

The number is stable against the transport, moving by $0.02\sigma$ across the photonuclear ensemble, because the energy shift, the reinterpreted slope, and the rate scaling cancel. It is stable against the detector too, and swapping our whole IceCube forward model for the Gaussian flux constraint of Ref.~\cite{Li:2025tqf} moves it by $0.04\sigma$. The published values~\cite{Li:2025tqf, Palmisano:2025abd, KM3NeT:2025ccp} differ from ours and from each other in how they treat the event's energy and KM3NeT's non-observation above it. Appendix~\ref{app:tension} goes through those differences one at a time. Note, that when using a down-scaled version of the trigger level effective area shown in Figure~\ref{fig:aeff}, the tension drops from $\sim 2.8\sigma$ to $\sim 2.2\sigma$ under the broken power law and from $2.2\sigma$ to $1.6\sigma$ under the single power law.

%%%%%%%%%%%%%%%%%%%%%%%%%%%%%%%%%%%%%%%%%%%%%%%%%%%%%%
%%%%%%%%%%%%%%%%%%%%%%%%%%%%%%%%%%%%%%%%%%%%%%%%%%%%%%

%%%%%%%%%%%%%%%%%%%%%%%%%%%%%%%%%%%%%%%%%%%%%%%%%%%%%%%%%%%%%%%%%%%%%%%%%%%%%%%%%%%%%%%%%%%
%%%%%%%%%%%%%%%%%%%%%%%%%%%%%%%%%%%%%%%%%%%%%%%%%%%%%%%%%%%%%%%%%%%%%%%%%%%%%%%%%%%%%%%%%%%

\section{Conclusion}
\label{sec:conclusions}

We have shown that the transport of a high-energy muon can be solved analytically once the fraction it loses in a collision does not depend on its energy. We describe the whole loss history with a single function, $\Phi$, and the muon range follows from its first two derivatives. It reproduces Monte Carlo propagation to within $3\%$ with nothing fitted, while the textbook range runs $10$ to $25\%$ long.

This calculation never reconstructs the loss spectrum. Each observable depends on the spectrum only through a few integrals, the mean energy on one and the muon range on a second, and any spectrum that shares those integrals gives the same answer. The loss coefficient $b_\mu$ is the first of those integrals. Today their uncertainty comes mostly from photonuclear scattering, so a more precise measurement at these energies would improve this approach.

With the muon range in hand, the detector's effective area is a one-line estimate, and with two instrument numbers per site, we reproduce four published areas within $1\%$. The same response gives point-source sensitivities at four sites, the energy of one event, and a consistency check between detectors, each without a simulation.

The approach extends beyond these examples. Power laws keep their shape in the transport, so a source of any spectral index propagates by a single factor, and a broken power law, a cutoff, or a monochromatic line is a sum of such power laws. Each of these used to need its own simulation. The same holds for a single particle, since $\Phi$ is linear in the loss spectrum. A stau~\cite{Meighen-Berger:2020eun, Bhattacharya:2025mdp} or a millicharged~\cite{VanDriessche:2019gph, ArguellesDelgado:2021lek, Xu:2022cvd} particle with energy-scale-invariant losses gets its own $\Phi$ and its own range from its loss spectrum alone. A new detector is then a change of two numbers, a new source a change of one index, and a new particle a change of one spectrum.

\section*{Acknowledgements}
We thank Sudipta Das, Diksha Garg, Mary Hall Reno, and Matheus Hostert for the helpful discussions and comments. We acknowledge the use of Claude, Anthropic's AI assistant for assistance in structuring and refining the prose of this paper.

\section*{Data and Code Availability}
The code behind this paper is Softpaws, which with the scripts that generate every figure and table is public at \url{https://github.com/MeighenBergerS/softpaws}. The IceCube data and response functions are the IceTracks-DR2 release~\cite{Abbasi:2026ehs}, and the PROPOSAL tables used for the loss spectrum are shipped with the code.

%%%%%%%%%%%%%%%%%%%%%%%%%%%%%%%%%%%%%%%%%%%%%%%%%%%%%%%%%%%%%%%%%%%%%%%%%%%%%%%%%%%%%%%%%%%
%%%%%%%%%%%%%%%%%%%%%%%%%%%%%%%%%%%%%%%%%%%%%%%%%%%%%%%%%%%%%%%%%%%%%%%%%%%%%%%%%%%%%%%%%%%

\appendix

%%%%%%%%%%%%%%%%%%%%%%%%%%%%%%%%%%%%%%%%%%%%%%%%%%%%%%%%%%%%%%%%%%%%%%%%%%%%%%%%%%%%%%%%%%%
\section{The Transport}
\label{app:diag}

Section~\ref{sec:exponent} started from the solution. We built the statistics of $W$ from the random walk it performs and checked in Section~\ref{sec:transporteq} that they satisfy the transport equation. Here we go the other way and solve Eq.~(\ref{eq:transport}) directly for a single muon, in three steps: we find the eigenfunctions of the collision operator, expand the starting distribution in them, and let each one evolve. We then derive the shape of $\Phi$ and the form of the source.

Under Eq.~(\ref{eq:scaleinv}) the collision operator does not change when every energy is rescaled by the same factor, so its eigenfunctions are power laws. We insert a power law $\phi = E^{-1-s}$ into Eq.~(\ref{eq:collision}). The gain term evaluates it at $E_y = E/(1-y)$, which returns $E^{-1-s}(1-y)^{1+s}$, and the Jacobian $1/(1-y)$ cancels one power, leaving
\begin{eqnarray}
  \Cqed\!\left[ E^{-1-s} \right]
  & = & - E^{-1-s} \int_0^1 \dd y \, \frac{\dd\Gamma}{\dd y}
        \left[ 1 - (1-y)^s \right] \nonumber \\
  & = & - \Phi(s) \, E^{-1-s},
\end{eqnarray}
which is Eq.~(\ref{eq:eigen}). Each power law therefore evolves on its own, as $E^{-1-s} e^{-\ell\Phi(s)}$.

The starting distribution of a single muon is a sum of such power laws. Along a line $\mathrm{Re}\, s = c$ inside the range where $\Phi$ converges, the delta function has the expansion
\begin{equation}
  \delta(E - E_0) = \frac{1}{2\pi i} \int_{c - i\infty}^{c + i\infty} \dd s \, \frac{E_0^s}{E^{1+s}},
  \label{eq:mellin}
\end{equation}
which is the inverse of the transform $\int \dd E \, E^s \phi$ that Section~\ref{sec:transporteq} takes. Letting each power law decay at its own rate solves Eq.~(\ref{eq:transport}) with $S = 0$
\begin{equation}
  \phi(\ell, E) = \frac{1}{2\pi i} \int_{c - i\infty}^{c + i\infty} \dd s \, \frac{E_0^s}{E^{1+s}} \, e^{-\ell\Phi(s)}.
  \label{eq:phisol}
\end{equation}
In the logarithmic loss $w = \ln(E_0/E)$, Eq.~(\ref{eq:phisol}) is the inverse Laplace transform of Eq.~(\ref{eq:genfn}), and it is the integral we evaluate numerically for Table~\ref{tab:losslaw}, along $c = 0$, where it becomes a Fourier integral. Taking its moments returns Eq.~(\ref{eq:momentlaw}). The single muon and a power-law spectrum thus share $\Phi$, because the single-muon solution is a superposition of the spectral modes.

The shape of $\Phi$ follows from the random walk of Section~\ref{sec:loglossW}. The accumulated logarithmic loss $W$ of Eq.~(\ref{eq:Wdef}) only grows, and its steps are independent and drawn from the same spectrum. In the language of probability theory, Eq.~(\ref{eq:genfn}) makes $\Phi$ the Laplace exponent of $W$, and every such exponent is a Bernstein function~\cite{Bertoin:1996levy, Schilling:2012bern}, whose derivatives alternate in sign. That gives $\Phi(0) = 0$, $\Phi' > 0$ and $\Phi'' < 0$.

Finally, the weak vertex acts on the flux the way a collision acts on a muon. The source of Eq.~(\ref{eq:transport}), which Section~\ref{sec:transporteq} set aside, is the flux folded with the charged-current cross section and its inelasticity distribution $P(y_w)$~\cite{Gandhi:1998ri, IceCube:2018pgc}
\begin{equation}
    S(E_0) = n_N \int_0^1 \frac{\dd y_w}{1-y_w} \, P(y_w) \,
    \left[ \phi_\nu \sigma_{\nu N} \right]\!\left( \frac{E_0}{1-y_w} \right),
    \label{eq:source}
\end{equation}
where $n_N$ is the nucleon number density. The vertex rescales the energy by $(1-y_w)$, as a collision does, so a power law of index $A$ passes through it with only a numerical factor
\begin{equation}
    S(E_0) = n_N \, I(A) \, \sigma_{\nu N}(E_0) \, \phi_\nu^{\oplus}(E_0) \propto E_0^{-1-A},
    \label{eq:sourcemode}
\end{equation}
where $\phi_\nu^{\oplus}$ is the neutrino flux arriving at Earth and $I(A) = \langle (1-y_w)^A \rangle_P$ is close to $0.8$ for the indices of interest. A power-law source therefore excites the single eigenfunction $E^{-1-A}$.

The photonuclear cross section rises slowly with energy~\cite{Bezrukov:1979hj, Abramowicz:1997ms, Butkevich:2001aw, Dutta:2000hh}, so the real loss spectrum carries a slow extra power $E^\beta$. Acting on a power law, such a loss spectrum still returns a power law, with its index shifted by $\beta$ and weighted by $\Phi_0$ at the shifted index, where $\Phi_0$ is the $\Phi$ of the energy-scale-invariant loss spectrum. A single muon instead sees its loss spectrum change along the track, so we apply the energy-scale-invariant results at the muon's current energy, as Appendix~\ref{app:renewal} does for the muon range.

\subsection{The Drift-Diffusion Limit}
\label{app:fp}
\label{app:moments}

The drift-diffusion treatment of Refs.~\cite{Palmisano:2025abd, Palmisano:2026sid} is the quadratic truncation of $\Phi$, and in the language of Section~\ref{sec:loglossW} it is a Gaussian logarithmic loss. Written as $A(b_\mu + d_\mu/2) - A^2 d_\mu/2$, Eq.~(\ref{eq:km}) is the generating function of a $W$ with mean rate $b_\mu + d_\mu/2$ and variance rate $d_\mu$. A Gaussian lets $W$ decrease, so no positive loss spectrum has this $\Phi$, and it turns negative above
\begin{equation}
    A_{\mathrm{die}} = 1 + \frac{2 b_\mu}{d_\mu},
    \label{eq:die}
\end{equation}
which for water at $1$~PeV is $A_{\mathrm{die}} = 9.3$. Cutting Eq.~(\ref{eq:km}) after its second term gives the column depth over which a spectrum of index $A$ falls by $1/e$
\begin{equation}
    L_{\mathrm{DD}} = \frac{1}{A b_\mu - \frac{A(A-1)}{2} d_\mu},
    \label{eq:LDD}
\end{equation}
which is the soft volume of Ref.~\cite{Palmisano:2026sid}. That work adds the collisions above a cut $y_{\mathrm{cut}}$ as corrections to the soft result, and its first correction is $-\Phi_{\mathrm{hard}}/\Phi_{\mathrm{soft}}$ times it. That correction is the first term in the expansion of $1/\Phi = 1/(\Phi_{\mathrm{soft}} + \Phi_{\mathrm{hard}})$ in powers of $\Phi_{\mathrm{hard}}$, and Eq.~(\ref{eq:Phi}) sums the whole series, with no cut.

\subsection{The Muon Range}
\label{app:renewal}

The muon range is a first-passage problem. We ask at what column depth the accumulated logarithmic loss $W$ of Eq.~(\ref{eq:Wdef}) first exceeds $w = \ln(E_0 / E_{\mathrm{thr}})$, the logarithmic energy the muon can shed before it falls below threshold. Since $W$ only grows, the muon is still above threshold at depth $\ell$ as long as $W(\ell) < w$, and the mean crossing depth is the integral of that probability
\begin{equation}
  L(E_0) = \int_0^\infty \dd\ell \, \mathbb{P}\!\left[ W(\ell) < w \right].
  \label{eq:Ldef}
\end{equation}
To evaluate it we use the potential density $u(w)$ of Section~\ref{sec:energy}, the column depth a muon spends at each value of the accumulated loss, $u(w) = \int_0^\infty \dd\ell \, p_\ell(w)$ with $p_\ell$ the distribution of $W(\ell)$. Its Laplace transform follows from Eq.~(\ref{eq:genfn})
\begin{equation}
  \int_0^\infty \dd w \, e^{-sw} \, u(w) = \int_0^\infty \dd\ell \, e^{-\ell \Phi(s)} = \frac{1}{\Phi(s)}.
  \label{eq:potential}
\end{equation}
The range is the cumulative column $\int_0^w u$, whose transform is $1/(s\Phi(s))$. Expanding that transform about $s = 0$ and inverting term by term gives Eq.~(\ref{eq:Lclosed}), which is the renewal theorem of probability theory with its second-moment correction~\cite{Feller:1971prob}.

The two numbers the muon range needs, $\Phi'(0)$ and $-\Phi''(0)$ of Eq.~(\ref{eq:Phimoments}), are the mean and the variance of the logarithmic loss per unit column depth. Both weight the hardest collisions, where $-\ln(1-y)$ diverges, and there the family of Eq.~(\ref{eq:betafam}) is $8\%$ low on $\Phi'(0)$ and $56\%$ low on $-\Phi''(0)$. We therefore integrate PROPOSAL's tabulated spectrum directly.

Over a descent of several decades the loss spectrum changes, and the muon range can follow it. Each order of magnitude in energy costs a column depth $\ln 10/\Phi'(0; E)$ at the energy where the muon is, while the overshoot belongs to the threshold it crosses, so
\begin{equation}
  L(E_0) = \int_{\ln E_{\mathrm{thr}}}^{\ln E_0} \frac{\dd \ln E}{\Phi'(0; E)}
  - \frac{\Phi''(0; E_{\mathrm{thr}})}{2 \, \Phi'(0; E_{\mathrm{thr}})^2}.
  \label{eq:Lrun}
\end{equation}
Holding the loss spectrum at the production energy returns Eq.~(\ref{eq:Lclosed}), which is short by $4\%$ at a parent energy of $10^5$~GeV and $11\%$ at $10^8$ (Table~\ref{tab:renewal}). We use the running form throughout. The estimate of Section~\ref{sec:closedform} instead holds the loss spectrum at $100$~TeV, a decade above threshold, where a descending muon spends most of its column depth, and the muon range then stays within $6\%$ of the running one over the energies of the tables.

Ionization cannot go inside $\Phi$, since it removes a fixed \textit{amount} of energy per unit length where the construction needs a fixed \textit{fraction}. A $1$~TeV threshold sits within a factor of two of $E_c$, so the last factor of a few in energy is not radiative. We therefore splice at $E_* = 10$~TeV, following the radiative range down to $E_*$ and the mean-loss law of Eq.~(\ref{eq:dedx}) below it
\begin{equation}
  L(E_0) = \frac{\ln(E_0/E_*)}{\Phi'(0)} - \frac{\Phi''(0)}{2\Phi'(0)^2}
  + \frac{1}{b_\mu(E_a)} \ln\frac{E_a + E_c}{E_{\mathrm{thr}} + E_c}.
  \label{eq:Lsplice}
\end{equation}
The muon jumps across $E_*$ as it jumps across any level, and on average it lands the mean overshoot below it, at $E_a = E_* e^{-\langle\delta\rangle} = 7.1$~TeV with $\langle\delta\rangle = -\Phi''(0)/2\Phi'(0)$ (Wald's identity), where we evaluate $b_\mu$. Moving $E_*$ by a factor of three either way moves the muon range by $3\%$, which is the systematic the splice carries.

An upgoing muon is born in rock and crosses only the ice or the water beneath the array before it is seen. Standard rock puts $\Phi'(0)$ per unit column depth $26\%$ above water at $1$~PeV through its $Z^2/A$ (Table~\ref{tab:inputs}). Since the muon only loses energy, its muon range splits at the energy $E_1$ it has when it leaves the rock
\begin{equation}
  L(E_0) = L_{\mathrm{rock}}(E_0 \to E_1) + X_{\mathrm{near}},
  \qquad L_{\mathrm{near}}(E_1 \to E_{\mathrm{thr}}) = X_{\mathrm{near}},
  \label{eq:Ltwomedium}
\end{equation}
where $X_{\mathrm{near}}$ is the column depth of ice or water between the rock and the detector, each segment is the running range of Eq.~(\ref{eq:Lrun}) in its own medium, and we solve for $E_1$ once per arrival direction. With $0.80$~km of ice below the center of IceCube and $0.40$~km of water below ARCA's, the muon range is about $0.83$ of its all-water value over the upgoing sky.

The spread of the range follows from the same transform, taken one order further
\begin{equation}
  \mathrm{Var}(L) = \frac{-\Phi''(0)\,w}{\Phi'(0)^3}
  - \frac{1}{\Phi'(0)^2}\left[ \frac{\Phi'''(0)}{3\Phi'(0)} - \frac{\Phi''(0)^2}{4\Phi'(0)^2} \right],
  \label{eq:Lvar}
\end{equation}
where $\Phi'''(0) = \langle -\ln^3(1-y)\rangle$ is a third quadrature of the same spectrum. The bracket is the variance of the overshoot. It enters with a minus sign, because a muon that overshoots further crossed sooner. Equation~(\ref{eq:Lvar}) holds to better than $2\%$ against a direct simulation of the same loss spectrum.

Both moments hold against PROPOSAL. We use the muon-range estimator of Ref.~\cite{Palmisano:2026sid} over production energies of $10^{5.5}$ to $10^{7.5}$~GeV, and stop the track at $100$~TeV so that it stays radiative. The mean muon range then agrees to $1.8\%$ and its spread to $3.5\%$. The second-order form returns the spread $19$ to $40\%$ low, because it carries $d_\mu = \langle y^2\rangle$ where the muon range needs $\langle \ln^2(1-y)\rangle$ (Table~\ref{tab:renewal}).

%%%%%%%%%%%%%%%%%%%%%%%%%%%%%%%%%%%%%%%%%%%%%%%%%%%%%%
\begin{table}[t]
\caption{The first-passage range of Eq.~(\ref{eq:Lclosed}) against the standard muon range of Eq.~(\ref{eq:range_standard}), for a muon in water with $E_{\mathrm{thr}} = 1$~TeV, with rates in km$^{-1}$ and ranges in km of water equivalent. The rates are integrals over the loss spectrum, with the angle brackets of Eq.~(\ref{eq:bdef}), and carry units of inverse column depth. Both ranges include the ionization splice of Eq.~(\ref{eq:Lsplice}).}
% Machine-written by scripts/2026_muon_transport/make_transport_table.py; do not edit by hand.
% Source: the softpaws library, PROPOSAL tabulation, water at 1.02 g/cm^3,
% E_thr = 1000 GeV. Ranges in km of water equivalent, rates in km^-1.
\begin{ruledtabular}
\begin{tabular}{lccccc}
$\log_{10}(E_\mu/\mathrm{GeV})$ & $4$ & $5$ & $6$ & $7$ & $8$ \\
\colrule
$b_\mu = \langle y \rangle$          & 0.342 & 0.363 & 0.380 & 0.400 & 0.426 \\
$\Phi'(0) = \langle -\ln(1{-}y)\rangle$ & 0.428 & 0.461 & 0.487 & 0.516 & 0.555 \\
$-\Phi''(0) = \langle \ln^2(1{-}y)\rangle$ & 0.293 & 0.345 & 0.379 & 0.412 & 0.453 \\
\colrule
$L$, Eq.~(\ref{eq:Lclosed}) & 5.73 & 10.76 & 15.62 & 20.21 & 24.52 \\
$R$, Eq.~(\ref{eq:range_standard})         & 5.73 & 12.10 & 18.30 & 24.21 & 29.79 \\
$L$, purely radiative       & 6.52 & 11.68 & 16.54 & 21.14 & 25.45 \\
\colrule
$L$, kernel frozen at production & 5.53 & 10.36 & 14.78 & 18.64 & 21.80 \\
\end{tabular}
\end{ruledtabular}

\label{tab:renewal}
\end{table}
%%%%%%%%%%%%%%%%%%%%%%%%%%%%%%%%%%%%%%%%%%%%%%%%%%%%%%

%%%%%%%%%%%%%%%%%%%%%%%%%%%%%%%%%%%%%%%%%%%%%%%%%%%%%%
\begin{table}[t]
\caption{Probability that a $1$~PeV muon has lost more than a factor $E_0/E$ of its energy over $1$~km of water, with $w = \ln(E_0/E)$. The Monte Carlo column is a PROPOSAL propagation of $4000$ muons~\cite{Koehne:2013gpa, Dunsch:2018nsc}, with statistical errors reaching $15\%$ in the last row. The next two invert Eq.~(\ref{eq:genfn}) numerically, with $\Phi$ from the tabulated spectrum and from the family of Eq.~(\ref{eq:betafam}), and the last is the Gaussian of the second-order expansion.}
\begin{ruledtabular}
\begin{tabular}{cccccc}
$w$ & $E_0/E$ & Monte Carlo & PROPOSAL & 3-moment & Gaussian \\
\colrule
$0.5$ & $1.6$  & $2.5\times10^{-1}$ & $2.5\times10^{-1}$ & $2.5\times10^{-1}$ & $4.0\times10^{-1}$ \\
$1.0$ & $2.7$  & $1.1\times10^{-1}$ & $1.1\times10^{-1}$ & $1.1\times10^{-1}$ & $2.9\times10^{-2}$ \\
$1.5$ & $4.5$  & $6.5\times10^{-2}$ & $6.0\times10^{-2}$ & $5.9\times10^{-2}$ & $2.0\times10^{-4}$ \\
$2.0$ & $7.4$  & $3.4\times10^{-2}$ & $3.5\times10^{-2}$ & $3.6\times10^{-2}$ & $1.0\times10^{-7}$ \\
$3.0$ & $20.1$ & $1.1\times10^{-2}$ & $1.3\times10^{-2}$ & $1.5\times10^{-2}$ & $9.6\times10^{-18}$ \\
\end{tabular}
\end{ruledtabular}
\label{tab:losslaw}
\end{table}
%%%%%%%%%%%%%%%%%%%%%%%%%%%%%%%%%%%%%%%%%%%%%%%%%%%%%%

%%%%%%%%%%%%%%%%%%%%%%%%%%%%%%%%%%%%%%%%%%%%%%%%%%%%%%%%%%%%%%%%%%%%%%%%%%%%%%%%%%%%%%%%%%%
\section{Instrument Inputs and the Response}
\label{app:inputs}

This appendix holds what a reader needs to evaluate the model at a site. Table~\ref{tab:inputs} collects the instrument numbers behind the comparisons of Section~\ref{sec:aeff}, Table~\ref{tab:recipe} the two fitted numbers per site, and Table~\ref{tab:estimate} every input of the estimate, Eq.~(\ref{eq:skyavg}), as Figure~\ref{fig:skyavg} uses it.

The published tables do not share a convention. The DR2 table is a $\nu_\mu$ area from $\nu_\mu$ simulation, while the ARCA21 bright-track curve is all-flavor and summed over $\nu + \bar\nu$ across $4\pi$, and mixing the two would cost a factor of order two. The two proposal-stage sites enter with their published geometry alone, P-ONE as seven clusters of $120$~m radius and $1$~km height at $2.16$~km depth~\cite{P-ONE:2020ljt}, and TRIDENT as one cylinder of $2$~km radius and $0.57$~km height at $3.10$~km~\cite{TRIDENT:2022hql}. We place both in sea water with the pure-absorption transmission their proposals assume.

\subsection{The Response Parameters}
\label{app:response}

We predict the reach of Eq.~(\ref{eq:reach}) from the light. We take a Frank-Tamm yield with the shower-track enhancement, attenuate it over the medium's own length, and collect it on the module photocathode. The visible radius then grows by $2.3$ effective attenuation lengths per order of magnitude in muon energy, so $\Lambda$ is one attenuation length. The effective length is the shorter of the absorption length and the diffusive length $\sqrt{\lambda_{\mathrm{abs}} \lambda_{\mathrm{scat}} / 3}$ at $400$~nm. The argument fixes the ordering and the scale of the reach, and it leaves each value uncertain by a factor of two across the photocathode's band. Deep South Pole ice absorbs at $110$ to $200$~m and scatters at $25$ to $50$~m~\cite{IceCube:2013llx}, which gives $30$ to $59$~m. Capo Passero water absorbs at about $50$~m~\cite{KM3Net:2016zxf, ANTARES:2004kfl}, which gives $41$ to $50$~m. The Cascadia Basin attenuates at $28$ to $35$~m~\cite{P-ONE:2020ljt}, which gives $24$ to $30$~m, and the South China Sea at $15$ to $27$~m~\cite{TRIDENT:2022hql}. These ranges are the predicted column of Table~\ref{tab:recipe}.

\begin{figure}[t]
\centering
\includegraphics[width=\columnwidth]{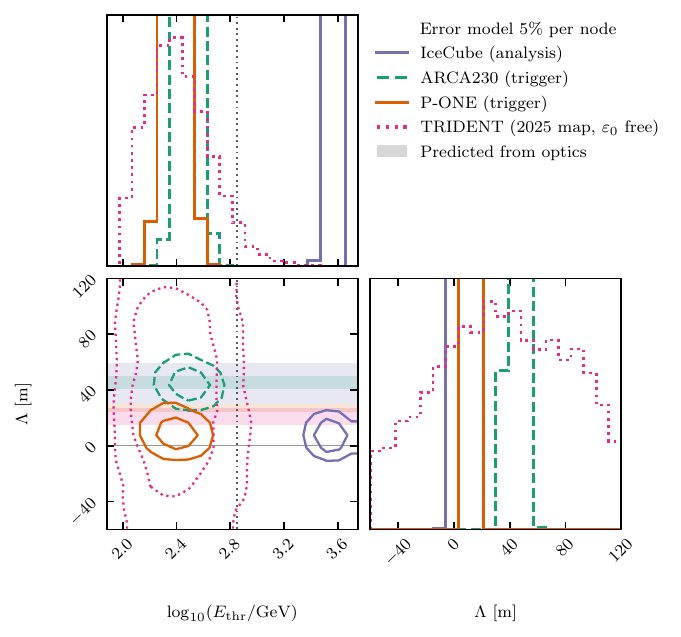}
\caption{The two instrument numbers of Eq.~(\ref{eq:aeff}) at the four sites, as $68\%$ and $95\%$ posterior contours with the physics fixed, against the reach each site's measured optics predict (bands, Appendix~\ref{app:response}). IceCube, ARCA230 and P-ONE carry an assumed $5\%$ error per tabulated point, and TRIDENT is fitted to the near-horizon cells of its 2025 map with its selection normalization free.}
\label{fig:aeff_posterior}
\end{figure}

Figure~\ref{fig:aeff_posterior} shows the two fitted numbers at each site against those predictions. The fitted reach lands in the predicted band at ARCA230, covers it within a wide interval at TRIDENT, and falls a factor of two below it at P-ONE, whose clusters are compact. At IceCube the analysis-level table returns $7$~m, since the quality selection removes the halo.

%%%%%%%%%%%%%%%%%%%%%%%%%%%%%%%%%%%%%%%%%%%%%%%%%%%%%%
%%%%%%%%%%%%%%%%%%%%%%%%%%%%%%%%%%%%%%%%%%%%%%%%%%%%%%

The fits hold the physics fixed because the tables cannot resolve it. A tabulated area reaches $\Phi$ only through the muon range of Eq.~(\ref{eq:Lclosed}), so scaling the loss spectrum by a constant factor scales Eq.~(\ref{eq:aeff}) by nearly its inverse, and the normalization undoes it. Freed alongside the threshold and the reach, that scale factor returns its prior at every site, and the cross-section slope trades against the reach. With the physics fixed the two numbers reproduce every table to $0.5$ to $1.3\%$ in the root mean square, against $0.3$ to $0.6\%$ with five parameters.

The selection normalization is set from outside, because a table does not determine it. At IceCube the normalization, the threshold and the reach lie on a flat degeneracy, $0.956$ with $3.5$~TeV and $7$~m or $0.75$ with $0.7$~TeV and $23$~m at the same deviance. We hold it at $0.956$, the value the hemisphere calibration of the first-principles optical chain returns, and the trigger-level sea tables at unity. None of the tables is released with errors, so we assume $5\%$ per tabulated point, independent and Gaussian in the logarithm, at which the deviances are $0.5$ to $3.6$ for $8$ to $12$ degrees of freedom.

We fit TRIDENT on the flat-selection cells of its 2025 map~\cite{Morton-Blake:2025vpp}, over the decade it shares with the 2022 table, with the normalization free. The fit returns a reach of $+38$~m, with a $68\%$ interval of $-19$ to $+97$~m, and a threshold of $248$~GeV. We do not use the 2022 sky average~\cite{TRIDENT:2022hql}, whose upgoing band above $1$~PeV falls a factor of $1.4$ below the model, a direction-dependent selection that the 2025 map makes explicit.

We build the loss-model budget of Section~\ref{sec:aeff} by swapping in each published parameterization of a loss channel, one at a time. Since $\Phi$ is linear in the loss spectrum, it carries each swap without approximation. Photonuclear scattering is the entire budget~\cite{Bezrukov:1979hj, Abramowicz:1997ms, Butkevich:2001aw, Dutta:2000hh}, and its effect shrinks on the way to an effective area, which averages the loss rate over the whole descent. It goes from $\pm 28\%$ on the mean loss at $10^8$~GeV to $\pm 2\%$ on the IceCube area at $10^5$~GeV and $\pm 8\%$ at $10^8$.

%%%%%%%%%%%%%%%%%%%%%%%%%%%%%%%%%%%%%%%%%%%%%%%%%%%%%%
\begin{table*}[t]
\caption{Inputs to the tabulated effective area of Section~\ref{sec:aeff}, for IceCube and for the two KM3NeT/ARCA configurations. The upper block is common to every site and the lower block changes between them, and no entry is fitted to an effective area.}
\begin{ruledtabular}
\begin{tabular}{llccc}
Quantity & Symbol & IceCube & ARCA230 & ARCA21 \\
\colrule
\multicolumn{5}{l}{\textit{Transport and interaction, common to all sites}} \\
Mean charged-current inelasticity~\cite{IceCube:2018pgc} & $\langle y_w \rangle$ & \multicolumn{3}{c}{$0.20$} \\
Mean neutral-current inelasticity~\cite{Gandhi:1995tf, Gandhi:1998ri} & $\langle y \rangle_{\mathrm{NC}}$ & \multicolumn{3}{c}{$0.25$} \\
Total-to-charged-current ratio~\cite{Gandhi:1998ri, Cooper-Sarkar:2011jtt} & $\sigma_{\mathrm{tot}}/\sigma_{\mathrm{CC}}$ & \multicolumn{3}{c}{$1.4$} \\
Tau branching to muons~\cite{ParticleDataGroup:2024cfk} & $B_{\tau\to\mu}$ & \multicolumn{3}{c}{$0.174$} \\
Mean muon energy fraction in $\tau \to \mu$~\cite{Dutta:2000hh, Bugaev:2003sw} & $\langle z \rangle$ & \multicolumn{3}{c}{$0.30$} \\
Mean $\nu_\tau$ energy fraction in $\tau$ decay~\cite{Dutta:2000hh, Bugaev:2003sw} & $\langle z_\nu \rangle$ & \multicolumn{3}{c}{$0.40$} \\
Neutrino-nucleon cross section & $\sigma_{\nu N}(E_\nu)$ & \multicolumn{3}{c}{BGR18~\cite{Bertone:2018dse}} \\
Muon loss coefficients & $b_\mu$, $d_\mu$ & \multicolumn{3}{c}{PROPOSAL~\cite{Koehne:2013gpa, Dunsch:2018nsc}} \\
Standard rock loss rate at $1$~PeV & $b_\mu$, $\Phi'(0)$ & \multicolumn{3}{c}{$0.481$, $0.615$~km$^{-1}$} \\
\colrule
\multicolumn{5}{l}{\textit{Instrument}, IceCube~\cite{IceCube:2016zyt} and ARCA~\cite{KM3Net:2016zxf, KM3NeT:2024paj}} \\
Instrumented shape & -- & hexagonal prism & two cylinders & one cylinder \\
Footprint radius, area equivalent & $R_{\mathrm{det}}$ [km] & $0.564$ & $0.517$ & $0.221$ \\
Instrumented height & $h$ [km] & $1.00$ & $0.632$ & $0.632$ \\
Instrumented volume & $V_{\mathrm{det}}$ [km$^3$] & $1.00$ & $1.06$ & $0.097$ \\
Projected area, range over zenith & $A_{\mathrm{proj}}$ [km$^2$] & $1.00$ to $1.55$ & $1.33$ to $2.13$ & $0.19$ to $0.32$ \\
Projected area, solid-angle mean & $\langle A_{\mathrm{proj}} \rangle$ [km$^2$] & $1.43$ & $1.87$ & $0.30$ \\
Depth of the instrumented volume & $d$ [km] & -- & $3.18$ & $3.18$ \\
Rock beneath & $h_{\mathrm{below}}$ [km] & $0.37$ & $0.08$ & $0.08$ \\
Available column, vertical & $X$ [g\,cm$^{-2}$] & Earth chord~\cite{Dziewonski:1981xy} & $3.25 \times 10^5$ & $3.25 \times 10^5$ \\
Solid angle of the tabulation & -- & $\sin\delta > 0$ & $4\pi$ & $4\pi$ \\
Flavor convention of the table & -- & $\nu_\mu$ & $\nu_\mu$ & all flavor, $\nu + \bar\nu$ \\
\end{tabular}
\end{ruledtabular}
\label{tab:inputs}
\end{table*}
%%%%%%%%%%%%%%%%%%%%%%%%%%%%%%%%%%%%%%%%%%%%%%%%%%%%%%
%%%%%%%%%%%%%%%%%%%%%%%%%%%%%%%%%%%%%%%%%%%%%%%%%%%%%%
\begin{table}[t]
\caption{Every input of Eq.~(\ref{eq:skyavg}) except for the cross section, for the four sites as Figure~\ref{fig:skyavg} uses them.}
% Machine-written by scripts/2026_muon_transport/make_estimate_table.py; do not edit by hand.
% Sources: example 89's first-principles inputs, the site geometry of the softpaws library, BGR18 with sigma_tot = 1.4 sigma_CC, PREM.
\begin{ruledtabular}
\begin{tabular}{lcccc}
 & IceCube & ARCA230 & P-ONE & TRIDENT \\
\colrule
$\epsilon_0$ & $0.956$ & $1$ & $1$ & $0.7$ \\
$E_{\mathrm{thr}}$ [GeV] & $3500$ & $300$ & $300$ & $300$ \\
$\Lambda$ [m] & $0$ & $46$ & $11$ & $21$ \\
$k$, Eq.~(\ref{eq:kreach}) & $0.00$ & $0.13$ & $0.10$ & $0.02$ \\
$\langle A_{\mathrm{proj}} \rangle$ [km$^2$] & $1.43$ & $1.87$ & $1.48$ & $8.07$ \\
$V_{\mathrm{det}}$ [km$^3$] & $1.00$ & $1.06$ & $0.32$ & $7.16$ \\
\colrule
$T$, $10^{5}$~GeV & $0.67$ & $0.66$ & $0.66$ & $0.66$ \\
$T$, $10^{6}$~GeV & $0.36$ & $0.35$ & $0.35$ & $0.35$ \\
$T$, $10^{7}$~GeV & $0.17$ & $0.16$ & $0.16$ & $0.16$ \\
\end{tabular}
\end{ruledtabular}

\label{tab:estimate}
\end{table}
%%%%%%%%%%%%%%%%%%%%%%%%%%%%%%%%%%%%%%%%%%%%%%%%%%%%%%
%%%%%%%%%%%%%%%%%%%%%%%%%%%%%%%%%%%%%%%%%%%%%%%%%%%%%%%%%%%%%%%%%%%%%%%%%%%%%%%%%%%%%%%%%%%

\subsection{The Analytic Recipe}
\label{app:recipe}

Before any sky average, the estimate at a single arrival direction $\Omega$ is
\begin{eqnarray}
  A_{\mathrm{eff}}(E_\nu, \Omega) &=& \epsilon_0 \, n_N \, \sigma_{\mathrm{CC}}(E_\nu) \, e^{-X(\Omega)/\Lambda_\nu(E_\nu)} \nonumber \\
  && \times \left[ A_{\mathrm{proj}}(\Omega, E_0) \, L(E_0) + V_{\mathrm{det}} \right],
  \label{eq:aeffcf}
\end{eqnarray}
where $L$ is the muon range of Eq.~(\ref{eq:skyavg}), held at one loss spectrum in water and capped at a sea site by the water above the detector. The projected area of the instrumented shape at a zenith angle $\theta$ is
\begin{equation}
  A_{\mathrm{proj}}(\theta, E_0) = \pi R_{\mathrm{eff}}^2 \, |\cos\theta| + c \, R_{\mathrm{eff}} \, h \, \sin\theta,
  \label{eq:aproj}
\end{equation}
where the first term is the footprint seen from above and the second the side seen from the horizon, $h$ is the instrumented height, and $c = 2$ for a cylinder and $2.1$ for IceCube's hexagonal prism. Only the radius grows with the reach of Eq.~(\ref{eq:reach}).

We work the recipe through for KM3NeT/ARCA230. The instrument is two cylinders of radius $R_{\mathrm{det}} = 0.517$~km and height $h = 0.632$~km, with $V_{\mathrm{det}} = 1.06$~km$^3$ at a depth of $3.18$~km, and we take $E_{\mathrm{thr}} = 313$~GeV, $\Lambda = 45$~m and the trigger-level normalization of unity. The transport enters through the two lengths of Eq.~(\ref{eq:skyavg}), $1/\Phi'(0) = 2.17$~km and $-\Phi''(0)/2\Phi'(0)^2 = 0.81$~km from the $100$~TeV column of Table~\ref{tab:renewal}. The cross section is BGR18~\cite{Bertone:2018dse} with $\sigma_{\mathrm{tot}} = 1.4 \, \sigma_{\mathrm{CC}}$, the nucleon density is $n_N = N_A \rho = 6.1 \times 10^{23}$~cm$^{-3}$ for sea water, with $N_A$ Avogadro's number and $\rho$ the density, and the column depth is $3.18~\mathrm{km} / \cos\theta$ of water above the horizon and the PREM chord~\cite{Dziewonski:1981xy} below it. The published curve is a whole-sky average, so we average too. Equation~(\ref{eq:skyavg}), with $\langle A_{\mathrm{proj}} \rangle = 1.87$~km$^2$ and $T$ taken over each hemisphere, returns $530$~m$^2$ at $1$~PeV. Averaging Eq.~(\ref{eq:aeffcf}) direction by direction returns $487$~m$^2$, and the full model returns $604$~m$^2$. The two averages differ because the first takes the projected area at its mean, while above a PeV the surviving directions hug the horizon and see the side of the instrument.

\subsection{Corrections to the Estimate}
\label{app:corrections}

Equation~(\ref{eq:aeff}) restores the three effects of Section~\ref{sec:theresponse}. The rock below the array enters through the two-medium muon range of Eq.~(\ref{eq:Ltwomedium}), while the Earth's regeneration and the tau channel need the constructions below.

\subsubsection{Earth Transmission with Regeneration}
\label{app:regeneration}

Treating the Earth as a pure absorber deletes neutrinos that scatter neutral-current. We keep them on a discrete ladder $E_k = E_\nu (1 - \langle y \rangle_{\mathrm{NC}})^k$, treating the neutral-current interaction as a fixed fractional loss, so each rung obeys
\begin{equation}
  \frac{\dd \varphi_k}{\dd X} = -N_A \sigma_{\mathrm{tot}}(E_k) \, \varphi_k + N_A \sigma_{\mathrm{NC}}(E_{k-1}) \, \varphi_{k-1},
  \label{eq:ladder}
\end{equation}
with $\varphi_k(0) = \delta_{k0}$ and $X$ the column depth traversed, taken from PREM along each chord~\cite{Dziewonski:1981xy}. Averaged over the upgoing hemisphere, regeneration raises the transmission by $27\%$ at $1$~PeV and $38\%$ between $10$ and $100$~PeV, which matches the full coupled cascade~\cite{Vincent:2017svp, Safa:2021ghs}. The tau channel regenerates as well, since the tau decays back to a neutrino~\cite{Halzen:1998be, Beacom:2001xn}. For an equal flavor ratio at Earth~\cite{Pakvasa:2007dc, Bustamante:2015waa} it carries the whole track rate at the nadir above $10^7$~GeV, where the $\nu_\mu$ transmission is $2 \times 10^{-4}$.

\subsubsection{Tau Transport}
\label{app:tau}

A tau neutrino adds tracks through $\nu_\tau \to \tau \to \mu$, and the rate from the two channels together is
\begin{eqnarray}
    \frac{\dd N}{\dd t \, \dd E \, \dd\Omega} &=& I(A) \, n_N \Big\{ \phi_{\nu_\mu}^{\oplus} \sigma_{\nu_\mu N} \left[ V_{\mathrm{det}} + V_{\mu}(E) \right] \nonumber \\
    && + \phi_{\nu_\tau}^{\oplus} \sigma_{\nu_\tau N} B_{\tau\to\mu} M(A) \, V_{\tau}(E) \Big\},
    \label{eq:totalrate}
\end{eqnarray}
where $B_{\tau\to\mu}$ is the tau branching fraction to muons~\cite{ParticleDataGroup:2024cfk} and $M(A) = \langle z^A \rangle$ is the moment of the muon's energy fraction $z = E_\mu / E_\tau$ over the polarized decay spectrum $g(z) = 2(1-z)^2(1+2z)$
\begin{equation}
    M(A) = \frac{4}{(A+1)(A+2)(A+3)} + \frac{8}{(A+2)(A+3)(A+4)},
    \label{eq:decaymoment}
\end{equation}
with $M(0) = 1$ and $M(1) = \langle z \rangle = 0.30$. Each channel has its own effective volume. Along an arrival direction with available column depth $x$, it attenuates the neutrino over the column depth $x - \ell$ it crosses before interacting and the lepton over the $\ell$ that remains. For the muon channel it is
\begin{equation}
    V_{\mu}(E) = \int_0^x \dd\ell \, A_{\mathrm{proj}}(E, \ell) \, e^{-\ell\Phi_\mu(A)} \, e^{-(x-\ell)/\Lambda_{\nu_\mu}},
    \label{eq:vsoftmu}
\end{equation}
and for the tau channel the muon's attenuation $e^{-\ell\Phi_\mu(A)}$ becomes $\mathcal{R}_\tau(\ell, A)$, the same attenuation along a two-stage path on which the tau travels, decays, and its muon covers the rest of the column
\begin{equation}
    V_{\tau}(E) = \int_0^x \dd\ell \, A_{\mathrm{proj}}(E, \ell) \, \mathcal{R}_\tau(\ell, A) \, e^{-(x-\ell)/\Lambda_{\nu_\tau}}.
    \label{eq:vsofttau}
\end{equation}
Every number in this paper takes the prompt limit, in which the tau decays where it is made and loses nothing on the way. Above $10^9$~GeV the decay length reaches the muon range and the two stages no longer separate, a regime we do not treat. Up to $10^8$~GeV the limit holds, because a tau radiates $(m_\mu/m_\tau)^2 \simeq 1/280$ as strongly as a muon and its decay length stays below $5$~km.

\section{Validation}
\label{app:declination}

Does the response hold direction by direction? Along the declination axis nothing is fitted. The two instrument numbers are fitted to the hemisphere average and neither depends on declination, so any variation from band to band tests the transport and the geometry. We compare the four upgoing bands of the DR2 table with two configurations. The static model uses the instrumented footprint, no reach and no normalization, so it has no fitted number at all. The calibrated model is the full model at the two numbers fitted to the hemisphere average. We leave out the downgoing hemisphere, whose atmospheric-muon self-veto the model does not carry~\cite{Schonert:2008is, Gaisser:2014bja}. We also leave out the deep-northern bins at $\sin\delta = 0.98$ above $10^{6.5}$~GeV, whose published floor no regeneration can reach in a table binned in the neutrino's true direction~\cite{Abbasi:2026ehs}.

%%%%%%%%%%%%%%%%%%%%%%%%%%%%%%%%%%%%%%%%%%%%%%%%%%%%%%
\begin{figure}[t]
\centering
\includegraphics[width=\columnwidth]{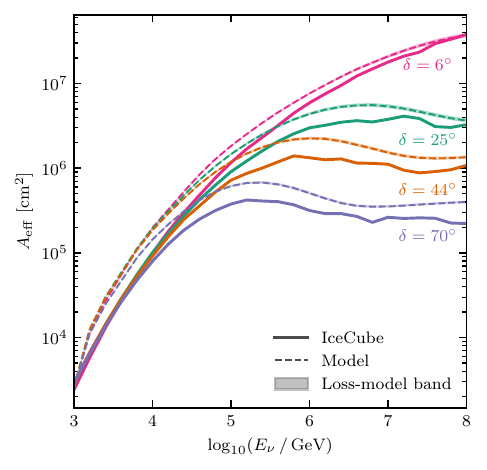}
\caption{Effective area in four upgoing declination bands of the DR2 table (solid)~\cite{Abbasi:2026ehs} compared to the first-principles model with the instrumented footprint and no normalization, evaluated band by band (dashed). Nothing is fitted, and the table sits at $0.69$ of the model over the hemisphere with a band-to-band spread of $10\%$.}
\label{fig:aeff_bands}
\end{figure}
%%%%%%%%%%%%%%%%%%%%%%%%%%%%%%%%%%%%%%%%%%%%%%%%%%%%%%

Figure~\ref{fig:aeff_bands} shows the static model. Over $10^5$ to $10^{7.8}$~GeV the published table sits at $0.69$ of it in every band and at every energy, with a spread of $10\%$ between bands.

The calibrated model meets the table at $0.85$ and $0.98$ in the two horizon bands and falls below it by factors of $1.3$ and $2.3$ in the next two. The shape therefore transfers in the horizon bands, which carry most of the hemisphere's acceptance, and degrades in the deepest bands, in opposite directions at IceCube and ARCA, which we attribute to the tables and their selections.

\subsection{Point-Source Ceiling Against IceCube}
\label{app:pointsource}

\begin{figure}[t]
\centering
\includegraphics[width=\columnwidth]{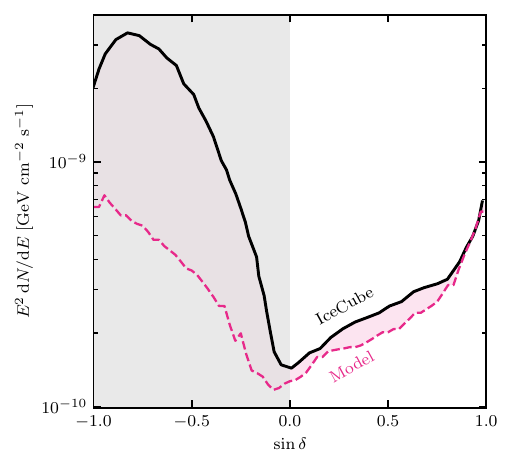}
\caption{The point-source limit of Eq.~(\ref{eq:pslim}) compared with the published fourteen-year track sensitivity of IceCube, with the model run at that analysis' own spectrum and exposure~\cite{IceCube:2025lev}.}
\label{fig:ps_sensitivity}
\end{figure}

Figure~\ref{fig:ps_sensitivity} compares Eq.~(\ref{eq:pslim}) with the fourteen-year all-sky track sensitivity of Ref.~\cite{IceCube:2025lev}, evaluated at that analysis' own spectrum and exposure. The two agree to $\mathcal{O}(20\%)$ across the northern sky, with a median ratio of $0.85$ and a worst deviation of $22\%$. The shape carries no fitted freedom, since the reach is fitted to a sky average. Across the southern sky the ratio falls to $0.22$, because a real southern search fights atmospheric muons, which we do not model. The two curves also come from different selections, DR2 and IceCube's point-source track sample, so the agreement checks the scale of the limit while the selection differs between them.

The four-site limits of Figure~\ref{fig:ps_sites} need only the zenith dependence of the response. The effective area depends on the arrival direction only through $\cos\theta$, so a source's declination enters through the fraction of a sidereal day it spends in each zenith band. We leave $E_{\min}$ free at each declination because an atmospheric spectrum falls faster than an astrophysical one. Footprint and depth set the level of each curve, and against IceCube the median limit is stronger by a factor of $4.6$ at TRIDENT, $2.1$ at ARCA230 and $1.5$ at P-ONE.

%%%%%%%%%%%%%%%%%%%%%%%%%%%%%%%%%%%%%%%%%%%%%%%%%%%%%%
%%%%%%%%%%%%%%%%%%%%%%%%%%%%%%%%%%%%%%%%%%%%%%%%%%%%%%

\subsection{The Event Sample in Declination}
\label{app:events}

\begin{figure}[t]
\centering
\includegraphics[width=\columnwidth]{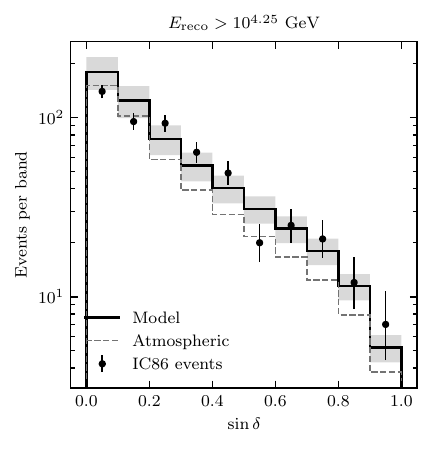}
\caption{Declination distribution of the upgoing IC86 sample over the prediction window, against the fixed-input prediction of Section~\ref{sec:events}. The dashed curve is the atmospheric component, the band the external-input uncertainty, and bands with zero observed events carry no point.}
\label{fig:events_dec}
\end{figure}

Figure~\ref{fig:events_dec} compares the fixed-input prediction of Section~\ref{sec:events} with the observed declination distribution over the whole prediction window. The model tracks all ten bands within the counting errors, and the astrophysical share grows toward the nadir, where the steep atmospheric flux dies fastest.

\begin{figure}[t]
\centering
\includegraphics[width=\columnwidth]{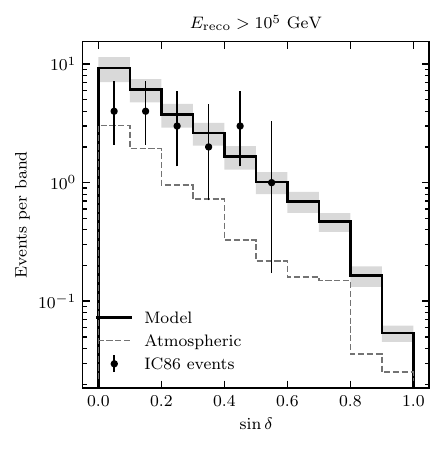}
\caption{As Figure~\ref{fig:events_dec}, above a reconstructed $10^5$~GeV, where the astrophysical component competes with the atmosphere.}
\label{fig:events_dec_high}
\end{figure}

Figure~\ref{fig:events_dec_high} repeats the comparison above a reconstructed $10^5$~GeV. There the model sits mildly high in the deepest bands, which is the deep-column behavior of Figure~\ref{fig:aeff_bands} seen at event level.

%%%%%%%%%%%%%%%%%%%%%%%%%%%%%%%%%%%%%%%%%%%%%%%%%%%%%%
%%%%%%%%%%%%%%%%%%%%%%%%%%%%%%%%%%%%%%%%%%%%%%%%%%%%%%

%%%%%%%%%%%%%%%%%%%%%%%%%%%%%%%%%%%%%%%%%%%%%%%%%%%%%%
%%%%%%%%%%%%%%%%%%%%%%%%%%%%%%%%%%%%%%%%%%%%%%%%%%%%%%

%%%%%%%%%%%%%%%%%%%%%%%%%%%%%%%%%%%%%%%%%%%%%%%%%%%%%%%%%%%%%%%%%%%%%%%%%%%%%%%%%%%%%%%%%%%
\section{The KM3-230213A Comparison}
\label{app:tension}

Why does our tension differ from the published ones? Table~\ref{tab:ladder} answers by going from the configuration of Ref.~\cite{Li:2025tqf} to ours one ingredient at a time, with every rung scored by the same likelihood ratio on two degrees of freedom. Their IceCube side is a Gaussian constraint on $(\phi_0, \gamma)$ at the collaboration's combined fit~\cite{Naab:2023xcz}. Their ARCA21 exposure is the $288$~days up to the event against our $335$, the full livetime of that configuration. The first rung already sits $1.5\sigma$ below their published number, a change of statistic from a Bayes factor over a marginalized posterior to a likelihood ratio that our machinery cannot reproduce. The middle of the ladder reaches the $1.9\sigma$ that the collaboration procedure returns. The last two rungs are our own choices, the per-event energy likelihood of Section~\ref{sec:energy}, built from the event's $3672$ triggered modules~\cite{KM3NeT:2025npi}, and IceCube's published broken power law. These two rungs take the statistic to the $2.8\sigma$ of Section~\ref{sec:tension}.

The gap between the two published numbers is not the IceCube detector response. Replacing their Gaussian by our entire IceCube forward model, with its response, exposure, threshold and a normalization three times lower, is worth $0.04\sigma$. The IceCube side enters only through where its region sits in $\log \phi_0$, and the ARCA side needs $\phi_0 \simeq 77$ against IceCube's ${\lesssim}2$, in units of $10^{-18}$~GeV$^{-1}$\,cm$^{-2}$\,s$^{-1}$\,sr$^{-1}$ per flavor. A factor of three at one end of a factor of forty does not show in the test statistic.

The comparison is most sensitive to the selection level at which the ARCA response is taken. The trigger-level curve runs a factor of $4$ to $6$ above the bright-track curve across this event's energy window~\cite{KM3NeT:2025npi}. Driving the same likelihood with it lowers the normalization one event requires from $\phi_0 \simeq 77$ to ${\simeq}17$, so the tension all but disappears. A trigger-level area suits a geometric bound, while inferring a flux needs the response that selected the event, which is the bright-track curve.

\begin{figure}[t]
\centering
\includegraphics[width=\columnwidth]{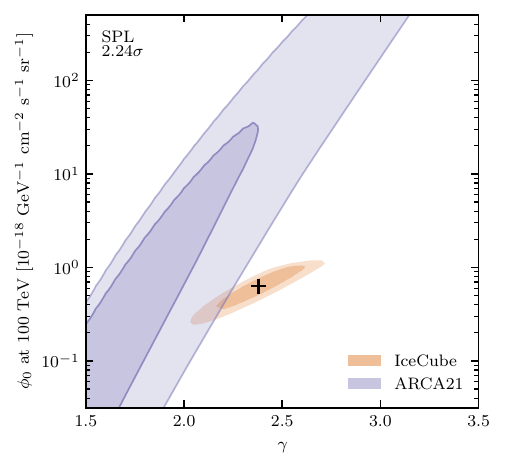}
\caption{The single-power-law companion of Figure~\ref{fig:flux_tension}, with the same construction: shaded regions at $68\%$ (darker) and $95\%$ (lighter), and the ARCA21 term combining the energy likelihood with the loss-model ensemble. The tension using a single power-law is $2.24\sigma$.}
\label{fig:flux_tension_spl}
\end{figure}

Figure~\ref{fig:flux_tension_spl} shows the single-power-law companion of Figure~\ref{fig:flux_tension}, where the two datasets are in tension at $2.24\sigma$.

%%%%%%%%%%%%%%%%%%%%%%%%%%%%%%%%%%%%%%%%%%%%%%%%%%%%%%
\begin{table}[t]
\caption{Path from the published $3.5\sigma$ of~\cite{Li:2025tqf} to the $2.75\sigma$ of Section~\ref{sec:tension}, each row changing one ingredient of the row above.}
\begin{ruledtabular}
\begin{tabular}{lcc}
Configuration & Significance & $\Delta$ \\
\colrule
Ref.~\cite{Li:2025tqf}, as published & $3.50\sigma$ & -- \\
\colrule
Their configuration, our statistic & $2.05\sigma$ & -- \\
$+$ our window, from $E_\mu$ & $2.30\sigma$ & $+0.25$ \\
$+$ the ARCA21 exposure, $335$~d & $2.24\sigma$ & $-0.06$ \\
$+$ the non-observation above it & $1.92\sigma$ & $-0.32$ \\
$+$ our IceCube forward model & $1.88\sigma$ & $-0.04$ \\
\colrule
$+$ the event's energy likelihood & $2.26\sigma$ & $+0.38$ \\
$+$ IceCube's broken power law & $2.75\sigma$ & $+0.49$ \\
\end{tabular}
\end{ruledtabular}
\label{tab:ladder}
\end{table}
%%%%%%%%%%%%%%%%%%%%%%%%%%%%%%%%%%%%%%%%%%%%%%%%%%%%%%

%%%%%%%%%%%%%%%%%%%%%%%%%%%%%%%%%%%%%%%%%%%%%%%%%%%%%%
%%%%%%%%%%%%%%%%%%%%%%%%%%%%%%%%%%%%%%%%%%%%%%%%%%%%%%

\clearpage
\bibliography{bibliography}

\end{document}